\documentclass[12pt,letterpaper,english,final]{article} 
\usepackage{amsfonts}
\usepackage[T1]{fontenc}
\usepackage{color}
\usepackage{pdflscape}
\usepackage{babel}
\usepackage{float}
\usepackage{mathtools}
\usepackage{caption}
\usepackage{url}
\usepackage{amsmath,bm,nccmath}
\usepackage{amsthm}
\usepackage{amssymb}
\usepackage{graphicx}
\usepackage{diagbox}

\usepackage{verbatim} 

\usepackage{natbib}
\usepackage[unicode=true,
 bookmarks=false,
 breaklinks=false,pdfborder={0 0 1},backref=false,colorlinks=true]{hyperref}
\usepackage[T1]{fontenc}

\usepackage[largesc]{newpxtext}
\usepackage{newpxmath}
\usepackage{titlesec}
\usepackage{titlesec,titling} 
\usepackage{slantsc}
\usepackage{cleveref}
\usepackage{etoolbox}

\usepackage{graphics,amsmath,amsthm,amssymb,amsfonts} 

\usepackage{mathrsfs}
\usepackage{mathtools}
\usepackage{multirow}
\usepackage[table]{xcolor}
\usepackage{graphicx}
\usepackage{tabularx}
\usepackage{color}
\usepackage{fancyvrb}
\usepackage{booktabs}
\usepackage{caption}
\usepackage{enumerate}
\usepackage{mathrsfs}
\usepackage{pdfpages}
\usepackage{setspace}
\usepackage{url}
\usepackage{lipsum}
\usepackage{physics}
\usepackage{tcolorbox}
\usepackage{tikz}
\usepackage[right=0.9in,top=0.9in,bottom=0.9in,left=0.9in]{geometry}
\usepackage{pgfplots}

\usepackage{epstopdf}
\usepackage{ragged2e} 
\usepackage{caption}

\usepackage{subcaption}
\usepackage{bbm}
\usepackage{epsfig}
\usepackage{threeparttable}
\usepackage{parskip}
\usepackage{cases}
\usepackage{hyperref}
\usepackage{color}

\usepackage[toc,page]{appendix}

\definecolor{darkblue}{rgb}{ 0.00,0.20,0.70}
\hypersetup{
 linkcolor=darkblue,
 citecolor=darkblue}
\makeatletter
\pdfpageheight\paperheight
\pdfpagewidth\paperwidth
\theoremstyle{plain}

\theoremstyle{plain}

\theoremstyle{plain} 

\theoremstyle{plain}

\usetikzlibrary{positioning}
\definecolor{sblue}{HTML}{4C72B0}
\definecolor{sorange}{HTML}{DD8452}
\definecolor{sgreen}{HTML}{55A868}
\definecolor{sred}{HTML}{C44E52}
\definecolor{mblue}{HTML}{23373b}
\definecolor{mred}{HTML}{b42828}
\definecolor{mgreen}{HTML}{D37418}
\pgfplotsset{compat = 1.10}
\usepgfplotslibrary{fillbetween}
\usetikzlibrary{patterns}
\def\munderbar#1{\underline{\sbox\tw@{$#1$}\dp\tw@\z@\box\tw@}}

\let\epsilon\varepsilon
\let\varepsilon\epsilon

\titleformat{\section}[block]{\Large\centering\scshape}{\thesection.}{0.5em}{}
\renewcommand\paragraph{\@startsection{paragraph}{4}{\z@}  {1.5ex \@plus1ex \@minus.2ex}  {-1em}  {\normalfont\normalsize\bfseries}}
\let\oldparagraph=\paragraph
\renewcommand\paragraph[1]{\oldparagraph{#1.}}
\spacing{1.3}
\makeatother
\providecommand{\corollaryname}{Corollary}
\providecommand{\lemmaname}{Lemma}

\providecommand{\propositionname}{Proposition}
\providecommand{\theoremname}{Theorem}

\renewenvironment{abstract}
 {\small
  \begin{center}
  \bfseries \abstractname\vspace{-.5em}\vspace{0pt}
  \end{center}
  \list{}{%
    \setlength{\leftmargin}{7mm}
    \setlength{\rightmargin}{\leftmargin}%
  }%
  \item\relax}
 {\endlist}

\allowdisplaybreaks

\def\equationautorefname~#1\null{((#1))\null}

\let\oldref\ref
\renewcommand{\ref}[1]{(\oldref{#1})}
\usepackage[bottom]{footmisc}

\titleformat{\part}[block]{\centering\fontsize{23}{23}\bfseries}{\thechapter.}{0.5em}{}

\renewcommand \thesection{\textbf{\arabic{section}}}

\begin{document}
\doublespacing
\onehalfspacing
\title{{\bf ChatGPT and Deepseek: Can They Predict the\\[1.2ex] Stock Market and Macroeconomy?}\thanks{\linespread{1.2}\selectfont We are grateful to Ilias Filippou, Guanhao Feng, Campbell R. Harvey, Jingyu He, Yi Huang, Lawrence J. Jin, Alejandro Lopez-Lira, Ziang Li, Jun Pan, Yuehua Tang, Dacheng Xiu, Xintong Zhan, Dexin Zhou, and seminar participants at City University of London, Dongbei University of Finance and Economics, Federal Reserve Bank of Atlanta, Hunan University, New York University (Shanghai), Imperial College, Peking University, Shanghai Advanced Institute of Finance,  Shanghai Jiatong University, Tsinghua University, University of Bath, University of Birmingham, University of Exeter, Washington University in St. Louis, Xiamen University, Xi’an Jiaotong-Liverpool University, York University, and Zhejiang University, and conference participants at 2024 Greater China Area Finance Conference and 2024 Summer Institute of Finance Conference. This paper, to include DeepSeek, supersedes a previous version circulated under the title "ChatGPT, Stock Market Predictability, and Links to the Macroeconomy". }\\ \vspace{0.8in}}

\author{Jian Chen\thanks{\linespread{1.2}\selectfont School of Economics \& Paula and Gregory Chow Institute for Studies in Economics, Xiamen University, China, 361005; e-mail: \protect\href{}{jchenl@xmu.edu.cn.}}
\and Guohao Tang\thanks{\linespread{1.2}\selectfont College of Finance and Statistics, Hunan University, China, 410006; e-mail: \protect\href{}{ghtang@hnu.edu.cn.}}
\and Guofu Zhou\thanks{\linespread{1.2}\selectfont \textbf{Corresponding author}: Olin School of Business, Washington University in St. Louis, St. Louis, Missouri, USA, 63130; e-mail: \protect\href{}{zhou@wustl.edu.}}		
\and Wu Zhu\thanks{Department of Finance, School of Economics and Management, Tsinghua University, China, 100084; e-mail: \protect\href{}{zhuwu@sem.tsinghua.edu.cn}}}
	
 \date{\vspace{0.5in} First Draft: July 2023 \\[1.2ex] Current Version: February 2025}
 \maketitle 

\thispagestyle{empty}

\clearpage\pagebreak



 \begin{center}
 \Large\textbf{ChatGPT and DeepSeek: Can They Predict the \\[1.2ex] Stock Market and Macroeconomy?}

 \end{center}
 
\sloppy

\vspace{0.3in}
\setstretch{1.5}

\begin{abstract}
\noindent
We study whether ChatGPT and DeepSeek can extract information from the Wall Street Journal to predict the stock market and the macroeconomy. We find that ChatGPT has predictive power. DeepSeek underperforms ChatGPT, which is trained more extensively in English. Other large language models also underperform. Consistent with financial theories, the predictability is driven by investors' underreaction to positive news, especially during periods of economic downturn and high information uncertainty. Negative news correlates with returns but lacks predictive value. At present, ChatGPT appears to be the only model capable of capturing economic news that links to the market risk premium.

\noindent
{\it JEL} classifications: C22, C53, G11, G12, G17 \\[1.2ex]
\end{abstract}

\thispagestyle{empty}

\restoregeometry
\pagebreak
\onehalfspacing
\setcounter{page}{1}
\setstretch{1.5}


\section{\textbf{Introduction}}
\setlength{\parindent}{2em}
\indent
Since its debut in November 2022, ChatGPT reached over 100 million users by January 2023. Sam Altman, one of its founders, has called
for a \$7 trillion investment—a sum that exceeds the US government’s revenue in 2023 and is enough to buy both Microsoft and Apple, the largest stocks in the US, combined.\footnote{Wall Street Journal, February 8, 2024.} Almost all academics and leaders in the industry and government recognize its revolutionary influence worldwide, making the understanding of its role in finance of great interest. \citet*{lopez2023can} is the first to study the predictive power of using ChatGPT on stock returns, whereas \citet*{jiang2023expected} and \citet{bybee2023ghost} examine more comprehensively various large language models. However, there is a lack of studies on predicting the  aggregated stock market, or equiavalently the market risk premium.  

In this paper, we present the first study on how ChatGPT may be utilized to forecast the stock market and investigate whether such predictability is both possible and significant. Our research question is of interest for three reasons. First, the predictability of the stock market is one of the central topics in finance as it tells us how the market risk premium varies over time which provides a benchmark for all investments in the economy.\footnote{ See, e.g., \citet{cochrane2008} for a detailed discussion and \citet{rapach2022asset} for a review of the literature.} Second, given that ChatGPT is so influential and powerful, it is of interest to know whether it can make a difference in addressing a key finance question. Third, our research answers whether \emph{Wall Street Journal} contains sufficient  information and, if so, whether it is fully incorporated into prices or not, and why. 

DeepSeek is a language understanding and generation model that serves as an alternative to ChatGPT. It has garnered enumerous attention recently due to its open-source nature, performance comparable to the most advanced GPT models in many aspects, and a notable reduction in training costs \cite{guo2025deepseek}.\footnote{ After its public release on January 27, 2025, Nvidia's stock price plummeted 17\%, triggering widespread global media and social media reactions. President Trump called it a 'wake-up call' for the US tech industry (see the Wall Street Journal on that date).}  While the two models are reportedly to have performed competitively in many ways, it may be noted that DeepSeek is trained primarily on Chinese and English datasets, with a strong emphasis on Chinese language processing. In contrast, ChatGPT has a  broader multilingual capabilities and is optimized for English-language interactions. Hence,   it is of great interest to see if ChatGPT can process the Wall Street Journal information better than DeepSeek. Therefore,  our study also compare the performance of the two and explains their differences in our context of financial forecasting.

In the era of big data, the amount of information produced has proliferated, increasing the complexity of information
processing. In recent decades, with the development of the natural language processing (NLP) technique, financial economists have begun to extract information about the stock market from various text sources such as the financial press \citep*[see a comprehensive
review by][]{loughran2020textual}. In our study, we focus on the news headlines and alerts on the front page of \emph{Wall Street Journal} from 1996 to 2022 and instruct ChatGPT-3.5 to identify good and bad news. We calculate the monthly ratio of good news to total news and bad news to total news. Using these two good and bad news ratios, we examine their return predictability on the aggregate stock market both in- and out-of-sample.

Empirically, we find that a ratio of good news is positively correlated with contemporaneous market returns, and it significantly predicts subsequent returns for the six months of the sample period from January 1996 to December 2022. The $R^2$ of the regression of one-month ahead market excess return on good news ratio ($NR^G$) is 1.37\%, with a slope of 0.53\%, statistically significant at the 5\% level. With the increase of prediction horizon, the $R^2$ rises and reaches 8.52\% over the annual horizon. The positive predictive power suggests that human investors might not efficiently capture good media news information. As a result, the information extracted by ChatGPT is incorporated into stock prices with a delay. In contrast, the negative news ratio has no forecasting power. This outcome aligns with expectations, showing a negative correlation with contemporaneous returns, indicative of investors' adeptness at quickly assimilating and responding to bad news, thereby precluding future market implications.

By contrast with ChatGPT, we find that Deepseek as well as the commonly used word lists proposed by \citet{loughran2011liability}  or the "small" pre-trained language model like Bert cannot predict the market. The weak performance of DeepSeek appears due to the way it was trained. Our empirical evidence suggests that the "word lists" method or Bert can not capture the context meaning of the words well. Specifically, in the "word lists" method, the meaning of each word is fixed and independent of the context (like the sentence or paragraph). In contrast, the "large" language models, endowed with hundreds of billions or more parameters, exhibit some "emergent abilities" not present in lists and BERT. For example, \cite{wei2022chain} shows that LLMs like GPT exhibit robust in-context learning under simple human instruction. We further find that the predictability of $NR^G$ remains robust even after including the common predictors like the lagged market returns and macroeconomic variables. These findings alleviate the concerns that the news content in \textit{Wall Street Journal} might already be encapsulated within economic fundamentals or that the return predictability comes from the continuation of stick prices.

Our results are supported by various economic explanations. First, they are consistent with models on investor attention \citep*[e.g.,][]{dellavigna2009investor,hirshleifer2009driven,ben2017depends, Anastassia2023}. Investors tend to pay more attention to bad news than good news, which is supported by Wall Street adage that “markets take the stairs up and the elevator down” and by index option trading where fund managers focus on hedging downside risk \citep*[e.g.,][]{chen2019demand}. Hence they react more quickly to so bad news, making it has little predictability. Second, our results align with theories on investment decision under information ambiguity. When the news quality is ambiguous, as evidenced with our news data of which ChatGPT identifies more than half as neutral, investors rationally react more to bad news than good news \citep*{epstein2008ambiguity}. Third, the predictability might be a result of information processing constraints and limits to arbitrage \citep*{lopez2023can}. Such information inefficiency is likely to be stronger in the aggregate stock market \citep*{xiao2022global}. Fourth, we show that the good news play, particularly, an importance role, especially when investors face downturns in the market, heightened economic policy uncertainty, or periods with more economically significant news than usual.

Our findings are robust in various settings. First, the predictability is robust to alternative prompts. We use keywords of "\emph{POSITIVE}" ("\emph{OPTIMISTIC}" or "\emph{GOOD}") and "\emph{NEGATIVE}" ("\emph{PESSIMISTIC}" or "\emph{BAD}") instead of "\emph{GOING UP}" and "\emph{GOING DOWN}" to ask ChatGPT-3.5. The newly defined $NR^G$ still predicts the market strongly. Second, we use ChatGPT-3.5 fine-tuning and ChatGPT-4, respectively, to identify good and bad news, and find that the return predictability remains significant and robust. Third, the strong return predictability of $NR^G$ also exists out-of-sample, which has become a critical assessment of predictability ever since \citet*{welch2008comprehensive}, because in-sample predictability can be unreliable due to parameter instability. Following the predictability literature, we evaluate \citet*{campbell2008predicting}'s out-of-sample $R_{OS}^2$ statistic and find that $NR^G$ still delivers statistically significant $R_{OS}^2$ of 1.17\% for the out-of-sample period from January 2006 to December 2022.

There is also significant economic value in the predictability of $NR^G$. Because of the significant positive $R_{OS}^2$, a mean-variance investor who allocates funds monthly between the market and risk-free assets can earn investment gains if the investor uses return forecasts based on $NR^G$ rather than using the historical average return. Indeed, the annualized certainty equivalent return (CER) gain is 4.92\% if the investor has a risk aversion degree of 3. This investment profit remains sizable after considering a proportional transaction cost of 50 basis points. The net-of-transaction-cost CER gains of $NR^G$ is 3.55\%. Moreover, the return forecasts of $NR^G$ generate a large annualized Sharpe ratio of 0.51, while the market has a Sharpe ratio of only 0.30. Our results are robust to alternative risk aversion coefficients, such as one or five.

A potential concern about out-of-sample studies on ChatGPT's predictability is a form of look-ahead bias. That is, some of the   forecasts are made by a model trained with future information. For example, when we identify a good news in January 2006, the starting of our out-of-sample period, with GPT-3.5 which is trained from a text corpus up until September 2021, it is unrealistic as
information in 2021 is not available in 2006. However, we argue that the if we were to use  a GPT that is trained with only data up to 2006, the result would be similar. There are three reasons for this. First, GPT-3.5 is designed to capture the meaning of the language for general purposes and does not use future marker return information to fine turn its performance. Since the meaning of   language overall is quite stable overtime. An evidence consistent this argument is that our results are similar even slightly weaker for GPT-4 which is trained with more and later data. Second, we compare the performance of GPT-3.5 against BERT, which was released in 2018 and so trained with data preceding 2018. We find that there is an almost constant and sustained outperformance of GPT-3.5 over BERT, with no evident performance spike during the 2018 to 2021 time frame, which suggests there is no substantial differences from access to   the newer information. Third, we also construct weekly good and bad news ratios, and examine whether there is a notable decline in performance subsequent to September 2021. Given that GPT-3.5 has no access to data post-September 2021, this is purely out-of-sample.
The performance should deteriorate if the extra textual information matters. However, we observe that the performance remains robust and there is no significant difference between before and after.

We further investigate the economic drivers of predictability. Our analysis reveals that a higher ratio of positive news signals an improving economic environment, while a higher ratio of negative news indicates potential economic decline. This result underscores ChatGPT's ability to extract relevant macroeconomic information from news texts, which subsequently impacts the overall stock market.


Last but not least, we explore the driving forces of predictability in several ways. First, human investors might only partially assimilate the textual information of good news during economic downturns. Related to this hypothesis, \citet{veronesi1999stock} illustrates how market responses to news can vary with the business cycle, notably showing an underreaction to good news in adverse economic times. ChatGPT potentially surpasses human analysis in interpreting news stories, enhancing its predictive strength during economic downturns. To shed light on this issue, using the Chicago Fed National Activity Index (CFNAI) as a proxy for economic conditions, we identify periods of high and low economic activity and find that it is indeed the case that \( NR^G \)'s predictive power is more pronounced during periods of low economic activity. Second, we hypothesize that information ambiguity, especially during times of high uncertainty, poses challenges for human interpretation as per \citet{zhang2006information}, whereas ChatGPT may exhibit superior capabilities in distilling contents. This hypothesis finds support in our regression analysis of future returns against interaction terms between \( NR^G \) and dummy variables based on economic policy uncertainty (EPU) proposed by \citet*{baker2016measuring}. Third, in line with the observations of \citet*{chan1996momentum}, investors might need more time to process newly released information fully. Hence, we propose that ChatGPT may excel in evaluating this novel information. Following \citet*{tetlock2011all}, we assess the novelty of information by examining the similarity of news stories to preceding relevant news. Our findings are consistent with the notion that \( NR^G \)'s return predictability is more substantial when the news is less similar to prior stories, thus supporting our hypothesis. All of the results collectively underscore ChatGPT's advanced ability to discern and leverage the predictive aspects of good news under varying economic conditions and information landscapes.

Recent developments in large language models (LLMs) have underscored the remarkable performance of DeepSeek‐R1, which, through reinforcement learning, substantially reduces computational burden while achieving performance on par with the most advanced GPT models \cite{guo2025deepseek}. A particularly compelling question is whether DeepSeek exhibits a comparable ability to extract valuable insights on financial markets as ChatGPT. Employing identical prompts and a dataset spanning from January 1996 to December 2022, we assess the effectiveness of DeepSeek‐R1 in predicting subsequent stock market returns and macroeconomic variables. Our analysis reveals that the positive and negative news signals derived from DeepSeek effectively capture contemporaneous stock market reactions, yet they lack forecasting power. A detailed examination further indicates that these news signals are highly correlated with the investor sentiment measure proposed by \cite{baker2006investor}, whereas sentiment metrics extracted via GPT do not display a statistically significant correlation with investor sentiment. Moreover, we document that the positive news extracted by DeepSeek fails to predict future macroeconomic fundamentals as captured by standard macroeconomic variables. Collectively, these findings suggest that DeepSeek primarily captures news signals reflective of investor sentiment, in contrast to ChatGPT, which appears more adept at uncovering information pertinent to macroeconomic fundamentals.


The paper contributes to the literature examining the role of the revolutionary language and artificial intelligence (AI) platform ChatGPT in extracting in-context financial information. While \citet*{lopez2023can}, \citet*{jiang2023expected}, and \citet{bybee2023ghost} are the major early studies in finance, there are various related studies in various fields, on which \citet*{Dong2023} provide a review. Different from those studies, \citet*{Guofu2024} use ChatGPT to examine unusual aspects of financial communications. In contrast to these studies about individual stock returns, we focus on ChatGPT's forecasting ability on the aggregate stock market. \citet*{bybee2023ghost} use ChatGPT to explore the correlation between expectations formed by professional forecasters and those generated by ChatGPT, finding that ChatGPT exhibits a similar tendency toward extrapolative expectations as professional forecasters. However, our study significantly diverges from his in two critical ways. First and foremost, our methods, primary findings, and underlying economic mechanisms are entirely distinct. Second, our measure of the ChatGPT information has strong out-of-sample predictive power, whereas his does not which has high correlation with CFO surveys shown to have no predictable power on the market by \citet*{he2023accurate}.

Our paper also contributes to the literature on textual analysis. In finance and accounting, textual analysis plays a pivotal role, scrutinizing text through the lenses of readability, similarity, and sentiment. Earlier studies are \citet*{antweiler2004all}, \citet*{tetlock2007giving}, \citet*{tetlock2008more}, \citet*{li2008annual}, and \citet*{tetlock2011all}, among others. Since \citet*{loughran2011liability}, their method of dictionary sentiment score has been widely used by the literature, e.g., \citet*{garcia2013sentiment}, \citet*{jiang2019manager}, and \citet*{cohen2020lazy}. Improving this method, the literature proposes unsupervised topic models \citep*{cong2019textual,bybee2021business,bybee2023narrative} and supervised models \citep*{manela2017news,garcia2023colour}. In contrast to these studies, we use large language models (LLMs), exemplified by ChatGPT, to extract the in-context information. Compared with the methods proposed by the prior studies, LLMs can capture both the syntax and semantics of text substantially better.

Our paper adds in a novel way to the large literature on market predictability. While there are many studies, such as \citet*{fama1977asset}, \citet*{campbell1988dividend} and \citet*{lo1988stock}, that support predictability, \citet*{welch2008comprehensive} argue for the lack of it on an out-of-sample basis. Subsequent studies, such as \citet*{campbell2008predicting}, \citet*{rapach2010out}, \citet*{henkel2011time} and \citet*{pettenuzzo2014forecasting}, find that the market is predictable even out-of-sample with innovative forecasting methods. \citet*{goyal2024comprehensive} recently raise doubts again on market predictability citing declining performance of predictors over time. In contrast with existing predictors. In this paper, we re-affirm the market predictability, which is the foundation of almost all modern asset pricing models with time-varying risk premia, with ChatGPT which captures unique information about the change political and social issues such as the COVID that traditional predictors fail to detect at least neither directly nor timely.

The remainder of the paper is organized as follows: Section \ref{sec: LLMs} briefly introduces the large language models and their backgrounds. Section \ref{sec: Methods} describes the empirical methods and data. Section \ref{sec: Results} provides our main empirical results. Section \ref{sec: eco_explanation} explores the economic mechanism of predictability. Section \ref{sec: conclusion}  concludes.
\section{\textbf{Large Language Models}}\label{sec: LLMs}
\setlength{\parindent}{2em}
\indent
This section briefly describes the background and the recent development of Natural Language Processing and Large Language Models (LLMs). The primary objective of NLP encompasses the development of robust and versatile representations for linguistic units such as words, sentences, paragraphs, and even larger text structures. These representations form the backbone for various downstream NLP applications, ranging from text classification and information extraction to entity recognition and question-answering systems. Recent advancements in this domain have been propelled by the advent of Large Language Models (LLMs), which leverage pre-trained language models or word embeddings to enhance the performance across diverse NLP tasks significantly \citep*{devlin2018bert,rogers2021primer}. We use a suite of state-of-the-art LLMs, including ChatGPT-3.5, ChatGPT-4 \citep{radford2019language}, BERT \citep{devlin2018bert}, and DeepSeek \cite{guo2025deepseek}\footnote{We also examine RoBERTa \citep{liu2019roberta} with results in the Online Appendix.}, 
to derive vector representations of news headlines. These models, with their distinct architectures and training methodologies, provide a comprehensive basis for analyzing and interpreting the semantic and syntactic nuances embedded in textual data.

Recent advancements in Large Language Models (LLMs) have introduced a paradigm shift in contrast with the paradigm of traditional word embedding, which assigns a static vector representation to each word in a predefined vocabulary irrespective of its contextual usage. These models employ attention mechanisms, a notable example being the Transformer architecture \citep{vaswani2017attention}, to dynamically learn word representations based on the surrounding context in which a word appears \citep{peters-etal-2018-deep}. This contextualized approach enables a more nuanced understanding of language, as the meaning of words can vary significantly depending on their usage in different sentences \citep{peters-etal-2018-deep,radford2019language}.

Here, we briefly describe two classes of LLMs-BERT and -ChatGPT and their applications in our context. Both models are built on the transformer \citep{vaswani2017attention}. Our empirical analysis make comparisons across models to examine how model size, robust adjustment, and fine-tuning in the context of financial documents affect the baseline results.

\subsection{\it\textbf{BERT}}
\setlength{\parindent}{2em}
\indent
Bidirectional Encoder Representations from Transformers (BERT) is a transformers model pre-trained on a large corpus of text data in a self-supervised way \citep*{devlin2018bert}.  The basic architecture of the BERT is a multi-layer bidirectional Transformer encoder based on the original implementation by \citet{vaswani2017attention}. Each transformer layer contains two sub-components: a multi-head self-attention mechanism and a fully connected feed-forward network. This design allows BERT to analyze and understand the context of a word based on all other words in a sentence, which is a significant departure from previous models that processed text unidirectionally. Using the attention mechanism, the embedding of each word depends on the context in which the word appears, in contrast to the traditional word embedding where each word is assigned as a fixed vector representation.  Moreover, the Transformer architecture supplants the sequential processing typical of recurrent neural networks (RNNs) with parallel processing attention mechanisms. This feature enables the model to assess the relevance of each word in a sentence without being constrained by their positional relationships.

BERT’s pre-training involves two distinct tasks: Masked Language Modeling (MLM) and Next Sentence Prediction (NSP). In MLM, the model randomly masks 15\% of the words in each input sentence and then runs the entire masked sentence through the model and tries to predict the masked words by processing the entire sentence. This task is designed to enable the model to learn the word representation based on the left and right context. In NSP task, the model concatenates two sentences as inputs during pre-training. Sometimes they correspond to sentences that were next to each other in the original text, sometimes not. The model then has to predict if the two sentences were following each other or not. Through this process, the model learns an inner representation (embedding) of words, sentences, and paragraphs, and the embeddings are then used to extract useful features for downstream analysis. 

The initial BERT model includes two versions: BERT BASE and BERT LARGE. Both take the same transformer structure but with different size of parameters. The BERT BASE consists of 110M while BERT LARGE includes 340M parameters. RoBERTa, developed by \citet{liu2019roberta}, emerges as an optimized iteration of BERT, enhancing its training methodology and dataset size to achieve improved performance across a range of NLP benchmarks. 


\subsection{\it\textbf{ChatGPT and DeepSeeek}}
\setlength{\parindent}{2em}
\indent
The Generative Pre-trained Transformer (GPT) series, developed by OpenAI, represents a significant milestone in NLP and artificial intelligence (AI). These models are a part of the broader family of Transformer-based models introduced by \citet{vaswani2017attention} and \citet{radford2019language}. Similar to the BERT, at its core, each GPT model is based on the Transformer architecture. Unlike traditional recurrent neural network (RNN) models that process input data sequentially, Transformers process all parts of the input data in parallel, significantly improving efficiency and scalability. This architecture allows GPT models to effectively capture the context dependencies of words and sentences. GPT models are characterized by their large-scale unsupervised pre-training. They are initially trained on vast amounts of text data, learning to predict the next word in a sentence. This pre-training phase enables the models to understand language patterns and structures deeply.

Compared to the BERT model, a striking feature of the ChatGPT is the significant scaling up in model parameters and training data. For example, ChatGPT-3 takes 175B parameters, GPT2 takes 1.5B parameters, but BERT only has 340M parameters. As the parameter scale of language models surpasses a threshold, a marked enhancement in performance becomes evident. Moreover, these advanced, large-scale models manifest distinctive capabilities, such as in-context learning, which remain absent in their smaller-scale counterparts. For example, ChatGPT-3 can solve few-shot tasks through in-context learning or sequential reasoning in complex tasks, while GPT-2 or BERT can not do that well. This delineation suggests that beyond quantitative improvements, qualitative transformations in model abilities emerge as a function of increased scale, which was documented as the \textit{emergent abilities} \citep{brown2020language,2022arXiv220607682W, LLMSurvey}. 

The concept of \textit{emergent abilities} in LLMs is rigorously delineated as capabilities absent in smaller models, such as BERT or GPT-2, but manifesting in more extensive models with hundreds or even thousands of billions of parameters \citep{2022arXiv220607682W, LLMSurvey}. These sources enumerate several \textit{emergent abilities}, including instruction tuning, in-context learning, and step-by-step reasoning. Instruction tuning is characterized by the LLMs' ability to adapt to new tasks guided solely by instructions, sans explicit examples, thereby exhibiting enhanced generalization. Within the ChatGPT series, this ability, also known as prompt engineering, forms the baseline of our empirical studies. In-context learning refers to the LLMs' capability to produce the anticipated output under natural language instructions or through a few task demonstrations without necessitating further training or gradient updates. Step-by-step reasoning, a concept highlighted in \citet{wei2022chain}, describes the LLMs' proficiency in navigating multi-step reasoning challenges using a chain-of-thoughts (C-o-T) approach.

DeepSeek-R1 distinguishes itself by leveraging a pure reinforcement learning approach that enables it to develop sophisticated reasoning capabilities without heavy reliance on extensive supervised fine-tuning \cite{guo2025deepseek}. Unlike conventional LLMs, which depend on large-scale supervised datasets, DeepSeek-R1 employs a multi-stage training pipeline—starting with cold-start data and followed by targeted reinforcement learning (RL) and fine-tuning. This approach allows it to naturally develop powerful chain-of-thought reasoning. This innovative strategy not only enhances efficiency by significantly reducing training costs but also delivers performance on par with state-of-the-art models, setting a new benchmark in reasoning tasks \cite{guo2025deepseek}.


In our empirical analysis, we initially apply instruction tuning to assess ChatGPT-3.5's capability in extracting pertinent information for stock market prediction and compare it with DeepSeek. Subsequently, we employ a range of fine-tuning techniques and instructions to evaluate the robustness of our baseline findings.

\section{\textbf{Empirical Methods and Data}}\label{sec: Methods}
\subsection{\it\textbf{Prompt}}
\setlength{\parindent}{2em}
\indent

Our baseline is to take advantage of the instruction ability of ChatGPT-3.5 to extract useful information for the stock market from media content. Specifically, news headlines are directly inputted into the model via a carefully designed prompt. This prompt functions as a directive mechanism, channeling the model's focus towards generating stock market predictions. In the operational framework of GPT models, a prompt is not merely a query or instruction posed to the model; it encompasses a broader scope, often integrating contextual information, specific input data, or illustrative examples to guide the model's response more effectively \citep{brown2020language}. This approach is predicated on the hypothesis that ChatGPT-3.5's advanced language processing and pattern recognition capabilities can discern and extrapolate relevant financial information and indicators from the textual data presented in news headlines.

We input the News headlines and alerts from \emph{Wall Street Journal} from 1996 to 2022 and instruct ChatGPT-3.5 (our baseline) to identify the good and bad news for the stock market. The prompt is:

"\textit{Forget all previous instructions. You are now a financial expert giving investment advice. I'll give you a news headline, and you need to answer whether this headline suggests the U.S. stock prices are GOING UP or GOING DOWN. Please choose only one option from GOING UP, GOING DOWN, UNKNOWN, and do not provide any additional responses.}"

As such, we count the number of good and bad news items each month and compute the percentage number (the number of good or bad news items divided by the total number of news items within a month). We use this percentage number to predict stock market returns. 

Compared to ChatGPT-3.5, ChatGPT-4 significantly expands the model size and offers greater accuracy and reliability across various language tasks like prediction and text understanding. The OpenAI finds that ChatGPT-4 can deal more with complex, multifaceted scenarios, which could be the case in finance, economics, and stock market analysis. To make a comparison, we use the same prompt as in ChatGPT-3.5 to judge the direction of stock movement based on the news headlines.

\subsection{\it\textbf{Few Shots}}
\setlength{\parindent}{2em}
\indent
In our baseline, we implemented a "zero-shot" prompting technique with ChatGPT-3.5, where the model was prompted to render judgments without access to any representative news headlines or their corresponding classifications. This method exemplifies zero-shot learning, in which large language models have shown remarkable adeptness. Nevertheless, it is observed that their efficacy often declines when applied to more intricate tasks under zero-shot settings \citep{kaplan2020scaling,brown2020language,touvron2023llama}. One can use the "few-shot" prompting method to amplify the model's capability for in-context learning. This approach integrates specific examples within the prompt, facilitating improved model performance.

In our study, one challenge arises from the fact that over 80\% of news headlines are typically categorized as \textit{UNKNOWN}. In a balanced few-shot scenario, at least ten examples are required to guarantee the presence of the \textit{GOING UP} and \textit{GOING DOWN} categories. However, incorporating numerous examples in a "few-shot" setup could lead to prohibitive token consumption per prompt. To circumvent this issue, we opted for direct model fine-tuning, tailoring the model parameters to better align with our specific research objectives.

\subsection{\it\textbf{Fine-Tuning}} \label{fine_tuning}
\setlength{\parindent}{2em}
\indent
In our study, a primary challenge in employing few-shot prompts for more precise classification is the potential mismatch in example distribution between the prompt and the actual data. Notably, \textit{UNKNOWN} classifications account for over 80\% of news headlines. Implementing a few-shot prompt with balanced classes (comprising two instances each of \textit{GOING UP}, \textit{GOING DOWN}, and \textit{UNKNOWN}) may inadvertently bias the model away from the "unknown" category.

As OpenAI has elucidated, 'fine-tuning surpasses few-shot learning by training on a more extensive set of examples than what is feasible within a prompt, thereby enabling the model to excel in a broader array of tasks.' Building on this insight, we explore whether fine-tuning ChatGPT-3.5 enhances model performance. To this end, we selected a random subset of 300 news headlines, manually and carefully classified each into \textit{GOING UP}, \textit{GOING DOWN}, and \textit{UNKNOWN}, and used this subset for fine-tuning ChatGPT-3.5's parameters.

In contrast to GPT models, BERT-type models do not adhere to a question-answer or completion format. To assess the predictive capabilities of BERT-type models in stock market analysis, we directly fine-tuned the BERT and ROBERTA models using the same 300 manually labeled news headlines. We divided this dataset into training (80\%) and validation (20\%) segments. The selection of hyperparameters—including early stopping (epochs), dropout rate, and parameter shrinkage—was guided by validation accuracy.

\subsection{\it\textbf{Embedding and News Similarity}}\label{similarity}
\setlength{\parindent}{2em}
\indent
In addition to the sentiment classification of news headlines, contemporary Large Language Models (LLMs) facilitate the derivation of vector representations for each headline. It significantly differs from traditional vector representation methods that predominantly rely on word frequency counts. Modern LLMs, incorporating sophisticated attention mechanisms, enable the incorporation of contextual information within sentence representations, thereby enriching the semantic understanding of the text. In our study, we employ a suite of advanced models - BERT \citep*{devlin2018bert}, RoBERTa \citep*{liu2019roberta}, ChatGPT-3.5, and ChatGPT-4 - for the extraction of vector representations from news headlines. These representations are subsequently utilized to compute the similarity of current news items relative to preceding ones, offering a nuanced approach to understanding temporal and thematic shifts in news content.

Specifically, we first average the vector embedding across all news headlines published within each month. For month $t$, we calculate the novelty of the news relative to the past five months as 
\begin{equation}
Novelty_{t} = 1 - \max_{1\leq j \leq 5} \text{Similarity}(\mathbf{e}_t,\mathbf{e}_{t-j})
\end{equation}
where $\mathbf{e}_t$ is the average vector representation of the news at month $t$, and \textbf{similarity} is the correlation similarity between two news.

\subsection{\it\textbf{Data Description}}
\setlength{\parindent}{2em}
\indent
We use front-page news on \emph{Wall Street Journal} which is available from Factiva. The news includes both headlines and business and finance alerts. The dataset employed in our study stands as one of the most expansive and comprehensive text corpora of business news investigated within the realm of economic literature to date. This dataset encompasses all articles published in front-page of \emph{Wall Street Journal} from January 1996 through December 2022, procured from the Dow Jones Historical News Archive. Notably, this dataset represents an unparalleled historical continuum of news articles available for purchase in digital format from Dow Jones \& Company, offering an extensive and unparalleled archival record of business news within the financial landscape.

The \emph{Wall Street Journal} stands as an iconic and invaluable resource within the domain of financial research \citep*{baker2016measuring,manela2017news,bybee2021business}, wielding a profound influence and playing a pivotal role in shaping the discourse and understanding of financial markets. As a venerable publication renowned for its comprehensive coverage of global financial markets, economic trends, corporate developments, and geopolitical events, the \emph{Wall Street Journal} serves as a cornerstone for academics, analysts, and practitioners alike seeking authoritative and real-time information.

\vspace{0.15in}
\centerline{[Insert Table \ref{Tab: summary} about here]}
\vspace{0.05in}

Table \ref{Tab: summary} presents the summary statistics of the \emph{Wall Street Journal} news dataset from January 1996 to December 2022. The dataset encompasses the total count of monthly news articles, categorized into "bad", "neutral", and "good" news based on their anticipated impact on the stock market—signifying the market is going down, uncertain, or going up —according to the ChatGPT-3.5 model's assessment in Panel A and DeepSeek-R1's assessment in Panel B. The average monthly news volume is 260.91, with a standard deviation of 116.12, indicating significant time volatility. A negative skewness of $-$0.22 suggests a mild leftward tilt in the distribution, while the median of 288 reflects the data's central tendency. The total news range extends from a minimum of 47 to a maximum of 577 items per month, encompassing 84,535 articles over the observed period.

Specifically, in Panel A, the "bad" news segment reports an average of 32.76 articles per month, with a lower variability (standard deviation of 23.93) than the total news count. This category exhibits a positive skewness of 1.15 and a median of 32, indicating a moderate rightward skew in its distribution. The monthly range for "bad" news varies from 0 to 142, totaling 10,613 articles.
The "neutral" news, forming the bulk of the dataset, averages 181.75 monthly articles with a standard deviation 74.34. This category shows an almost symmetrical distribution (skewness of $-$0.07) and a median of 190. The range for "neutral" news stretches from 43 to 422 monthly articles, amounting to 58,888. Lastly, the "good" news category, denoting positive market sentiments, maintains an average of 46.40 monthly articles with a standard deviation 28.25. It exhibits a slight positive skewness (0.10) and a median of 47. The monthly range for "good" news spans from 0 to 121, aggregating 15,034 articles.

Panel B shows the statistic summary of DeepSeek. Interestingly, both DeepSeek and ChatGPT have roughly the same number of neutral news. However, DeepSeek identifies more bad news, but fewer good news than ChatGPT.

\begin{figure}[H] 
\captionsetup{singlelinecheck=false,justification=raggedright,labelfont=bf}
\caption{\textbf{Time Series Plots of News Ratios}}\vspace{-0.2cm}
{\small This figure depicts the time series of monthly good news ratio ($NR^G$) and bad news ratio ($NR^B$) from January 1996 to December 2022. $NR^{G}$ ($NR^B$) is defined as the monthly proportion of good (bad) news identified by ChatGPT-3.5. It answers whether the input news means \textit{GOING UP} (\textit{GOING DOWN}) for stock market. The vertical bars correspond to the NBER-dated recessions. }\label{Fig: Time-series}

\begin{center}
\includegraphics[viewport=1.5cm 1.5cm 20cm 23cm]{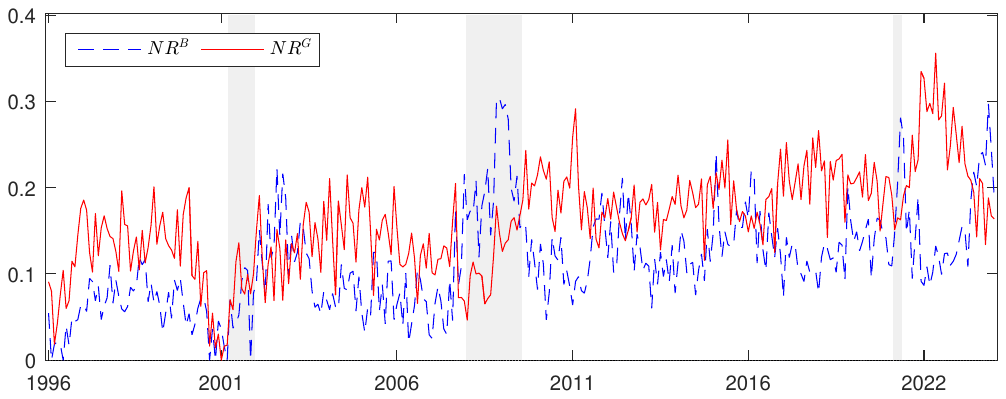}
\end{center}

\end{figure}

\vspace{0.15in}
\centerline{[Insert Figure \ref{Fig: Time-series} about here]}
\vspace{0.05in}

We define the monthly news ratio as the proportion of good news to the total monthly news count, denoted as \( NR^G \), and the proportion of bad news to the total monthly news count, represented by \( NR^B \). Figure \ref{Fig: Time-series} illustrates the temporal progression of these ratios, \( NR^G \) and \( NR^B \), through a time series plot. The figure shows that ChatGPT captures well some of the important episodes of the economy. Let us focus on $NR^G$, the good news ratio.  It has the largest decline during year 2000 to 2001, corresponding the burst of the dot.com bubble. The next major decline is around 2008, the housing bubble. There is a general decline around year 2000-2001 due to COVID. The later spike appears to signal the success of vaccine and then the good news stabilizes. In short, with a simple visual  analysis, the ChatGPT seems performs what it is designed for.

\section{\textbf{Empirical Results}}\label{sec: Results}
\subsection{\it \textbf{Baseline Regression Model}}
\setlength{\parindent}{2em}
\indent
This section explores the impact of the textual information captured by ChatGPT-3.5 on the aggregate stock market. We use a univariate regression model as follows:
\begin{equation}\label{Pred_Z}
R_{t+h} = \alpha + \beta\, NR_{t}^{K} + \epsilon_{t+h} \ , \qquad K=B\ \mbox{or}\ G \ ,
\end{equation}
where $R_{t}$ denotes the current market excess return on the S\&P 500 index at time $t$ for "$h=0$". This setting aligns with a contemporaneous regression framework. For scenarios where $h>0$, $R_{t+h}$ represents the average excess returns of the market portfolio from $t+1$ to $t+h$ (with $h$ being 1, 3, 6, 9, and 12 months), transitioning the equation into a predictive regression. Utilizing ChatGPT-3.5's zero-shot prompt capability, we identify instances of good or bad news from the front-page headlines of the \emph{Wall Street Journal}. The bad news ratio, $NR^{B}$, is quantified as the monthly proportion of bad news, while $NR^{G}$ represents the proportion of good news. The in-sample predictability of both $NR^{B}$ and $NR^{G}$ is examined by estimating the regression for the period from January 1996 to December 2022. A critical aspect of our analysis involves assessing the coefficient $\beta$ ($\hat\beta$) in the regression. The null hypothesis presumes that the ChatGPT-estimated textual information lacks predictive power, implying $\beta=0$ and reducing the regression to a model of constant expected return ($R_{t+h}=\alpha+\epsilon_{t+h}$). The alternative hypothesis, however, posits that $\beta$ is non-zero, indicating that GPT-extracted textual information holds significant predictive power for $R_{t+h}$. For the computation of the corresponding $t$-statistic for $\hat\beta$, we employ the \citet*{hodrick1992dividend} standard error.\footnote{In analyzing predictability over extended horizons, the \citet*{newey1986simple} may lead to over-rejection in finite samples, particularly when persistent regressors interact with serially correlated errors. The Hodrick's standard error addresses this issue by adopting the moving-average structure of the aggregated error, thereby offering enhanced performance \citep*{ang2007stock}.}

\vspace{0.15in}
\centerline{[Insert Table \ref{Tab: In-sample} about here]}
\vspace{0.05in}

Panel A of Table \ref{Tab: In-sample} presents the estimation results of regression  (\ref{Pred_Z}). We observe the slope coefficient for the contemporaneous regression on \( NR^{B} \) (when \( h=0 \)) at $-1.03\%$, with a \citet*{hodrick1992dividend} \( t \)-statistic of $-$2.70. However, coefficients across various predictive horizons are not statistically significant. In contrast, results for $NR^{G}$ reveal a more gradual market response. Here, the regression coefficient for the contemporaneous relationship between stock market returns and \( NR^{G} \) is 1.02\% and is statistically significant. This finding implies that ChatGPT-identified good news is positively correlated with the stock market. Notably, this significance persists with coefficients ranging from 0.51\% to 0.56\% over predictive horizons of 1 to 6 months, though it declines after the 9-month mark. Moreover, the predictive regressions exhibit substantial \( R^2 \) values, between 1.37\% and 8.52\%, which increase with longer horizons, highlighting the significant predictive power of \( NR^{G} \) for stock market returns. These results indicate that ChatGPT is capable of extracting aspects of good news not typically discerned by investors, leading to a delayed incorporation of this information into stock prices. The robustness of our findings is further supported by the \citet*{newey1986simple} \( t \)-statistic, as presented in Table IA 1 of the Internet Appendix.

Our findings presented in Table \ref{Tab: In-sample} notably diverge from the results of \citet*{tetlock2007giving}. Tetlock's study highlighted the predictive power of daily news pessimism on stock market trends, revealing a tendency for reversal within a five-day period, while suggesting a lack of predictive strength in news optimism. This divergence in outcomes is primarily attributable to the distinct methodologies adopted in textual analysis. \citet*{tetlock2007giving} employed the Harvard psychology dictionary in conjunction with word count frequencies for extracting text. This conventional technique assigns a static representation to each word, disregarding the context in which it is used. In stark contrast, our approach harnesses the potential of Large Language Models (LLMs), particularly their underlying transformer architecture. This modern method enables a more nuanced and contextually aware extraction of meanings at both the word and sentence levels \citep{vaswani2017attention,peters-etal-2018-deep}, providing a deeper insight into the semantic intricacies of financial news. In subsequent sections, we delve further into this comparison. We juxtapose our LLM-based approach with the word list methodology proposed by \citet*{loughran2011liability}, which focuses on the identification of positive and negative words. This comparative analysis is designed to shed light on the strengths and limitations of each method in the context of financial textual analysis.

Furthermore, the research conducted by \citet{frank2018does}, which leveraged firm-level data, posits that stock prices are prone to overreact to positive news and underreact to negative news. Conversely, our investigation reveals that the market's incorporation of positive news is characteristically gradual, in stark contrast to the immediate reaction elicited by negative news. This differentiation is consistent with the theoretical exposition by \citet{epstein2008ambiguity}, which posits that ambiguity-averse investors, when faced with news of uncertain quality, tend to adopt a worst-case perspective regarding its veracity. Consequently, this predisposes them to exhibit more pronounced responses to negative news compared to positive news, highlighting a fundamental asymmetry in news assimilation dynamics. Our finding rules out the possible interpretation of investor inattention, such as those advocated by \citet*[e.g.,][]{dellavigna2009investor,hirshleifer2009driven,ben2017depends}, which typically suggest an absence of immediate stock return responses to new information. A thorough examination of this divergence, along with an exploration of potential economic rationales underpinning these market behaviors, is undertaken in Section \ref{sec: eco_explanation}.

DeepSeek has similar contemporaneous slopes to ChatGPT, as showed by Panel B, the slope coefficients are statistically significant for both good and bad news ratios. However, these estimates become insignificant when predicting subsequent returns. This indicates that the information identified by DeepSeek is immediately incorporated into stock prices, and it does not have predictive power on futre stock returns. For this reason, we will not discuss DeepSeek in our further analysis on stock market predictability below. But we do compare DeepSeek and ChatGPT later on how they predict the macro economy.

In summary, our analysis reveals that a high proportion of good news, as identified by ChatGPT-3.5, exhibits a significant correlation with both the current and subsequent returns of the market portfolio, extending up to six months. Conversely, the ratio of bad news demonstrates a positive correlation exclusively with current returns, lacking predictive power for future returns. These findings underscore a distinct market response pattern: while the assimilation of positive news identified by ChatGPT-3.5 into market prices is gradual, the response to negative news is immediate. In contrast, both good and bad news identified by DeepSeek are incorporated into the market immediately, without any delayed responses.

\subsection{\it \textbf{Comparisons with Other Methods of Textual Analysis}}
\setlength{\parindent}{2em}
\indent
Our study underscores the robust predictive capability of ChatGPT-3.5 in discerning textual information for future market returns. This efficacy prompts an exploration of whether alternative methods of textual analysis offer comparable predictive insights. A notable approach in this regard is the word lists or bag-of-words methodology, prominently advocated by \citet*{loughran2011liability}. They introduced a specialized dictionary of \emph{positive} and \emph{negative} words, specifically designed for financial texts, an approach that has garnered widespread adoption, as evidenced in studies like \citet*{garcia2013sentiment}, \citet*{jiang2019manager}, and \citet*{cohen2020lazy}.

Employing this dictionary, we categorized front-page news from the \emph{Wall Street Journal} into "good" and "bad" news segments and subsequently computed their respective news ratios. We then re-estimated the regression (\ref{Pred_Z}), using the news ratio derived from this word list method as the regressor. The findings, as presented in Table \ref{Tab: comparison}, reveal a minimal impact of good news on stock market returns. In contrast, the ratio of bad news exhibits a significant correlation with current market returns but lacks the ability to predict future returns.

\vspace{0.15in}
\centerline{[Insert Table \ref{Tab: comparison} about here]}
\vspace{0.05in}

Our results align with the observations of \citet{tetlock2007giving}, particularly in underscoring the greater influence of news pessimism compared to optimism on stock returns. However, our findings diverge concerning the forecasting ability of bad news. \citet*{tetlock2007giving} posited a significant predictive power stemming from media pessimism, a claim not entirely supported by our analysis. A possible reason for this discrepancy could be attributed to the diminished novelty of information extracted via word lists. Since the introduction of this method in 2011, its widespread adoption might have lessened its novelty and, consequently, its predictive value to human investors.

In our subsequent analysis, we evaluate the potential of other LLMs for their abilities to discern textual information and to predict stock market returns. Specifically, we utilized BERT and RoBERTa, developed by \emph{google}, to analyze "good" and "bad"news from the front-page news of the \emph{Wall Street Journal}. Unlike the GPT framework, these models do not readily accommodate natural language instructions (prompts) to differentiate between positive and negative news. To overcome this limitation, we adopted an alternative approach. We began by randomly selecting 300 news articles, then manually classified each into categories: \textit{GOING UP}, \textit{GOING DOWN}, or \textit{UNKNOWN}. Following this classification, we embarked on a process of fine-tuning the parameters of the BERT model. This fine-tuning involved retraining and updating the model's parameters, with a focus on selecting hyper-parameters that optimized accuracy on our validation set.

Table \ref{Tab: comparison} delineates the forecasting outcomes for both the bad and good news ratios as identified by the BERT model. It is observed that the textual information extracted by BERT exerts minimal influence on stock market returns, with a notable exception being the contemporaneous regression that is contingent upon the good news ratio. In a similar vein, the RoBERTa model, as reported in Table IA 2 of the Internet Appendix, demonstrates limited predictive capabilities. These findings align well with the recent scholarly discourse \citep{brown2020language,2022arXiv220607682W,wei2022chain, LLMSurvey}, which posits that larger language models, boasting hundreds of billions or more parameters, exhibit unique "emergent abilities" that are not typically found in smaller models such as BERT and RoBERTa.

\begin{figure}[H]
\captionsetup{singlelinecheck=false,justification=raggedright,labelfont=bf}
\caption{\textbf{Word Cloud Identified by Various Language Models}}\vspace{-0.2cm}
{\small This figure shows the word distribution for the good news and bad news identified by various methods. The two sub-figures of the first row shows word distribution for the good news and bad news, respectively, identified by the ChatGPT-3.5. Similarly, the second and third rows show the word cloud for the good news and bad news identified by the word lists proposed by \citet*{loughran2011liability} and the BERT model, respectively. }\label{Fig: Word Cloud}

\centering
\includegraphics[width=1\textwidth]{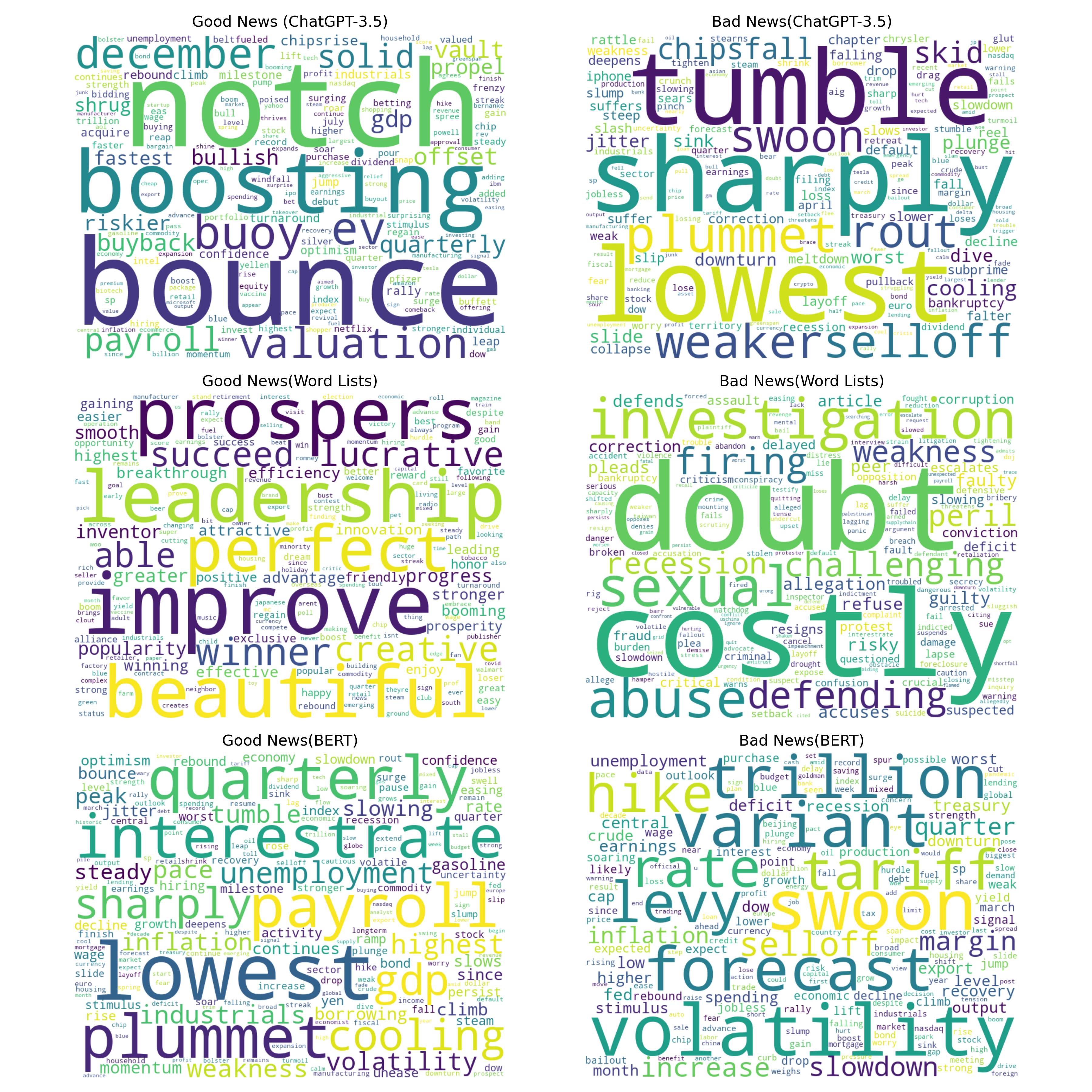}
\end{figure}

\vspace{0.15in}
\centerline{[Insert Figure \ref{Fig: Word Cloud} about here]}
\vspace{0.05in}

To elucidate the distribution of words within categories of the good news, bad news, and the unknown news as identified by ChatGPT-3.5, BERT, and the word lists method used in sentiment analysis \citep*{loughran2011liability}, we conducted a detailed word frequency analysis. This process involved several key steps:
\begin{itemize}
    \item First, we cleaned the news headlines by removing digits, punctuation, and stop-words using the NLTK package, a widely recognized tool in text analysis. This was followed by lemmatizing the text to retain the base form of each word.\footnote{For more details about NLTK, refer to \url{https://www.nltk.org/}.}
    \item Next, we computed the word frequency for each headline, subsequently aggregating these frequencies within each category: the good news, bad news, and the unknown news.
    \item Finally, we excluded the least frequent words and calculated the relative frequency of remaining words within each category.
\end{itemize}
Figure \ref{Fig: Word Cloud} illustrates the word clouds for good and bad news as classified by ChatGPT-3.5, the word lists approach of \cite{loughran2011liability}, and the BERT model. Notably, ChatGPT-3.5 adeptly captures context-sensitive words of the financial market. For instance, frequent terms in positive news include "bounce", "notch", "boosting", and "buoy" typically associated with favorable financial or economic conditions. Conversely, the word lists method yields words like "leadership", "beautiful", "improve", and "prospers" for positive news, which, despite being generally affirmative, are less commonly linked to financial contexts. Intriguingly, the BERT model appears less effective in this regard, incorrectly associating words such as "plummet" and "lowest" with positive news, thus contradicting economic intuition.

Overall, this section delves into a comparative analysis of text information predictability as extracted by various methodologies: ChatGPT-3.5, the conventional word lists approach, and other language models with relatively smaller parameter sizes, such as BERT and RoBERTa. Our investigation reveals that ChatGPT-3.5 demonstrates a notable "emergent" ability. It proficiently extracts significant stock market information that is not as effectively captured by either the traditional word lists approach or the smaller language models. This distinction highlights the advanced capability of ChatGPT-3.5 in processing and interpreting complex textual data relevant to financial markets.

\subsection{\it \textbf{Comparisons with Macroeconomic Predictors}}\label{Economic Variable}
\setlength{\parindent}{2em}
\indent
The content of \textit{Wall Street Journal} may already encompass key aspects of economic fundamentals, raising the question of whether the predictability of \( NR^{G} \) using ChatGPT-3.5 is merely reflective of these underlying economic variables. To explore this possibility, we extend our analysis by incorporating common economic variables as controls in our predictive model. The modified regression model is structured as follows:
\begin{equation} \label{Pred_Z_L}
R_{t+h}=\alpha + \beta\, NR^G_{t} + \psi\,\textbf{X}_{t} + \epsilon_{t+h} \ ,
\end{equation}
where \( R_{t} \) represents the current market excess return at time \( t \) when \( h=0 \). For instances where \( h>0 \), \( R_{t+h} \) denotes the average excess returns of the market portfolio from \( t+1 \) to \( t+h \), with \( h \) varying from 1 to 12 months. Thus, equation (\ref{Pred_Z_L}) functions as a predictive regression. Here, \( NR^{G} \) signifies the good news ratio as identified by ChatGPT-3.5, while \( \textbf{X}_t \) is a vector of 14 economic variables proposed by \citet{welch2008comprehensive}.\footnote{Data sourced from \url{https://sites.google.com/view/agoyal145}} A comprehensive description of these variables is available in \ref{appendix_a}.

Using all the macroeconomic variables together in a single regression may result in the potential collinearity issue. Instead, we opted to control for the first five principal components of these variables. The regression outcomes, detailed in Table IA 3 of the Internet Appendix, reveal that the coefficients of \( NR^G \) consistently maintain positive values and bear economic significance across various prediction horizons. These results are in alignment with the estimates reported in Table \ref{Tab: In-sample}, underscoring the robustness of our findings. Crucially, the statistical significance of \( NR^G \) remains after the inclusion of common economic variables as controls. This suggests that the information content of \( NR^G \) is not merely a reflection of these economic variables but rather provides distinct and valuable insights.

Additionally, we controlled for lagged market returns in our analysis. Given the significant correlation of \( NR^G \) with contemporaneous returns, it is imperative to ascertain whether the observed predictability of \( NR^G \) is inherently tied to the continuation of stock price. To alleviate this concern, we included the current market return as control variable in the predictive regressions based on \( NR^G \). The detailed results, presented in Table IA 4 of the Internet Appendix, show that the coefficient of \( NR^G \) retains its significance, resonating with the findings in Table \ref{Tab: In-sample}. Conversely, the coefficient for the current return demonstrates insignificance, suggesting that the predictability attributed to \( NR^G \) is not merely a manifestation of price momentum. In short, our analysis in this section underscores that the predictive power of \( NR^G \), as extracted by ChatGPT-3.5, is distinct and not merely an overlap with existing fundamental economic variables or stock market trend.

\subsection{\it \textbf{Robustness Check}}
\setlength{\parindent}{2em}
\indent
In this subsection, we show our results robust to alternative prompts, fine-tuning, and ChatGPT-4. We first use three alternative prompts:
\begin{enumerate}
\item 
"\textit{Forget all previous instructions. You are now a financial expert giving investment advice. I'll give you a news headline, and you need to answer whether this headline is PESSIMISTIC or OPTIMISTIC for the U.S. stock market. Please choose only one option from PESSIMISTIC, OPTIMISTIC, UNKNOWN, and do not provide any additional responses.}"
  
\item 
"\textit{Forget all previous instructions. You are now a financial expert giving investment advice. I'll give you a news headline, and you need to answer whether this headline is NEGATIVE or POSITIVE for the U.S. stock market. Please choose only one option from NEGATIVE, POSITIVE, UNKNOWN, and do not provide any additional responses.}"
  
\item 
"\textit{Forget all previous instructions. You are now a financial expert giving investment advice. I'll give you a news headline, and you need to answer whether this headline is GOOD or BAD for the U.S. stock market. Please choose only one option from GOOD, BAD, UNKNOWN, and do not provide any additional responses.}"
\end{enumerate}

\vspace{0.15in}
\centerline{[Insert Tables \ref{Tab: Prompts} and \ref{Tab: GPT4} about here]}
\vspace{0.05in}

Utilizing these three distinct prompts, we engaged ChatGPT-3.5 to analyze news articles and compute corresponding news ratios. The regression results based on these newly defined news ratios are tabulated in Table \ref{Tab: Prompts}, elucidating the outcomes of regression (\ref{Pred_Z}). We note that \emph{pessimistic} (or \emph{negative}) news appears to exert minimal impact on stock market returns, as evidenced in Panels A and C. In contrast, \emph{optimistic} (or \emph{positive}) news demonstrates a significant influence on market performance in Panels B and D. Specifically, in the contemporaneous regression, the slope coefficient for the \emph{optimistic} (or \emph{positive}) news ratio registers at 1.09\% (0.74\%) with a \citet*{hodrick1992dividend} \( t \)-statistic of 5.55 (3.31). This significance remains in the predictive regressions, with coefficients ranging from 0.43\% to 0.54\% in Panel B (and from 0.43\% to 0.51\% in Panel D). Table IA 5 of the Internet Appendix reports analogous results for the third prompt. These findings collectively attest to the robustness of ChatGPT-3.5's forecasting abilities across a variety of prompts.

Moreover, we employed both ChatGPT-3.5 fine-tuning and ChatGPT-4 to categorize front-page news from the \emph{Wall Street Journal} into good and bad news categories. The forecasting results obtained from these models are delineated in Table \ref{Tab: GPT4}. Generally, these outcomes are consistent with those detailed in Table \ref{Tab: In-sample}. Specifically, \( NR^B \), as determined through ChatGPT-3.5 fine-tuning, shows a significant correlation with current returns, yet lacks predictive power for future returns. In contrast, \( NR^G \) demonstrates a substantial influence on both current market returns and those over extended periods, with forecasting coefficients varying from 0.34\% to 0.80\%. This range highlights the pronounced predictive ability of \( NR^G \) for the stock market.

The findings for ChatGPT-4 align with the observed trend of immediate market reactions to both positive and negative news, as well as the gradual incorporation of positive news into stock prices, as shown in Table \ref{Tab: In-sample}. However, our analysis does not indicate a clear performance advantage of ChatGPT-4 over ChatGPT-3.5 in predicting stock returns. A closer investigation reveals that ChatGPT-4 demonstrates a tendency toward over-classification, where it is more likely than ChatGPT-3.5 to classify uncertain news as either positive or negative. This behavior reduces the predictability of positive news classifications by ChatGPT-4, as it potentially introduces noise into the forecasting process, leading to a slight shrinkage toward zero and reducing the statistical significance for the coefficient. Furthermore, ChatGPT-4's enhanced proficiency in handling multimodal data-integrating textual and visual elements-could dilute its focus on text-based financial predictions. This aligns with OpenAI's assertion that the model's training objective was not explicitly optimized for stock market forecasting but rather for understanding nuanced textual differences and multimodal capability. This broader focus may result in a trade-off between general language capabilities and domain-specific predictive accuracy. In fact, this is consistent with the stronger predictability of our fine-tuning model which aims at enhancing the stock predictability.

In general, these findings reinforce the robustness of ChatGPT-3.5’s forecasting ability, highlighting its adaptability and effectiveness in predicting stock returns without the added complexity of multimodal capabilities. This underscores that increased model sophistication does not necessarily translate into improved performance in financial prediction tasks, as evidenced by the limited predictive advantage of ChatGPT-4 over its predecessor.


\subsection{\it \textbf{Out-of-sample Performance}}
\setlength{\parindent}{2em}
\indent
This section is dedicated to assessing the out-of-sample return predictability of \( NR^B \) and \( NR^G \), as extracted through ChatGPT-3.5. While in-sample analysis facilitates more efficient estimation of parameters and thereby yields more precise return forecasts by leveraging the entire available data, studies such as \citet*{welch2008comprehensive} argue for the greater relevance of out-of-sample tests. These tests are deemed crucial in evaluating the actual predictability of returns in a real-time setting, providing a more authentic assessment of the model's predictive power in practical finance scenarios.

We initiate our out-of-sample forecast evaluation with a starting period from January 1996 to December 2005. This period serves as the basis for estimating the monthly predictive regression (\ref{Pred_Z}) using \( NR^B \) or \( NR^G \), thereby facilitating the generation of our first out-of-sample forecast in January 2006. The forecasted return is articulated as follows:
\begin{equation}\label{Pred_OS}
\widehat{R}_{t+1}= \widehat{\alpha}_t + \widehat{\beta}_t\ L^{I}_{t}\ ,
\end{equation}
where \( \widehat{\alpha}_t \) and \( \widehat{\beta}_t \) represent the ordinary least squares (OLS) estimates derived from regression (\ref{Pred_Z}). We subsequently engage in a recursive process, continually re-estimating regression (\ref{Pred_Z}) and constructing monthly out-of-sample forecasts in accordance with Equation (\ref{Pred_OS}). This approach is consistently applied to subsequent periods, extending up to the end of our sample period in December 2022. The selection of the initial in-sample estimation period was strategically made to ensure that the observations were ample for accurately estimating initial parameters, while also allowing for a sufficiently extended out-of-sample period for effective forecast evaluation.\footnote{\citet*{rossi2010shape} and \citet*{hansen2012choice} indicate that out-of-sample tests of predictive ability tend to exhibit improved size properties when the forecast evaluation period constitutes a relatively large proportion of the available sample, as is the case in our analysis.}

To assess the out-of-sample performance, we implement the widely used \citet*{campbell2008predicting}'s \( R_{OS}^2 \) and \citet*{clark2007approximately}'s \emph{MSFE-adjusted} statistical methods. The \( R_{OS}^2 \) is a measure of the proportional reduction in mean squared forecast error (MSFE) for the predictive regression forecast relative to a benchmark forecast. A positive \( R_{OS}^2 \) value indicates that the model forecast surpasses the benchmark in terms of MSFE. The benchmark in this context is the average excess return from the start of the sample period up to month \( t \), aligning with the constant expected excess return model delineated in Equation (\ref{Pred_Z}) with \( \beta=0 \). This implies that returns are not predictable, akin to the canonical random walk model with drift applied to stock prices. To determine whether the predictive regression forecast yields a statistically significant improvement in MSFE, we employ the \emph{MSFE-adjusted} statistic as proposed by \citet*{clark2007approximately}. This is used to test the null hypothesis \( H_0: R_{OS}^2 \leq 0 \) against the alternative hypothesis \( H_A: R_{OS}^2 > 0 \), which posits that the historical average MSFE exceeds that of the predictive regression forecast.

\vspace{0.15in}
\centerline{[Insert Table \ref{Tab: OOS} about here]}
\vspace{0.05in}

Table \ref{Tab: OOS} displays the out-of-sample forecasting results. We observe that while \( NR^B \) exhibits a negative \( R_{OS}^2 \), the \( R_{OS}^2 \) for \( NR^G \) stands at 1.17\% and is statistically significant according to \emph{MSFE-adjusted} statistics. A positive \( R_{OS}^2 \) suggests that the MSFE for out-of-sample return forecasts based on \( NR^G \) is significantly lower than the historical average, indicating substantial economic significance. Given the typically small \( R^2 \) in stock return predictions due to the high noise-to-signal ratio, the magnitude of \( R_{OS}^2 \) for \( NR^G \) is notably large. \citet*{campbell2008predicting} contend that a monthly \( R_{OS}^2 \) of 0.5\% can have significant economic implications, and our finding of an \( R_{OS}^2 \) over twice this threshold underscores its substantial economic relevance \citep*{kandel1996predictability}. In the following section, we will delve into the economic gains derived from this predictability.

Though the bad news ratio cannot predict the market alone, it might carry complimentary forecasting information. According to \citet*{rapach2010out}, the average combination of forecasts from individual economic variables surpasses the performance of a kitchen sink model, using all variables together in a single model, as suggested by \citet{welch2008comprehensive}. We find that the mean combination (MC) delivers a positive \( R_{OS}^2 \) of 0.39\%, better than the negative \( R_{OS}^2 \) of \( NR^B \). Furthermore, we use the iterated mean combination (IMC) and the iterated weighted combination (IWC) proposed by \citet*{lin2018forecasting}. The IMC forecast ($\hat R^{IMC}$) is,
\begin{equation}\label{IMC}
\hat R^{IMC}_{t+1|t} = (1-\hat \theta^{MC})\bar R_t + \hat \theta^{MC} \hat R^{MC}_{t+1|t} \ ,
\end{equation}
where $\bar R_t$ is the historical average forecast, $\hat R^{MC}_{t+1|t}$ is the mean combination forecast, and $\hat \theta^{MC}$ is the optimal weight from the first-order condition of the objective function:
\begin{equation*}
\theta=\frac{cov(R_{t+1}-\bar R_t,\,\hat R^{MC}_{t+1|t}-\bar R_t)}{var_t(\hat R^{MC}_{t+1|t}-\bar R_t)} \ . 
\end{equation*} 
The IWC forecast is,
\begin{equation}\label{IWC}
\hat R^{IWC}_{t+1|t} = (1-\hat \theta^{WC})\bar R_t + \hat \theta^{WC} \hat R^{WC}_{t+1|t} \ ,
\end{equation}
where $\hat R^{WC}_{t+1|t}$ is the weighted combination forecast suggested by \citet*{rapach2010out}. Table \ref{Tab: OOS} shows that both IMC and IWC generate positive and significant \( R_{OS}^2s \), 1.36\% and 2.51\%, respectively. The magnitude is larger than the \( R_{OS}^2 \) of using individual predictor \( NR^G \) or \( NR^B \).

For comparative analysis, we also examine the results for economic variables. Table \ref{Tab: OOS} report the mean combination forecasts of 14 economic variables of \citet{welch2008comprehensive}. The \( R_{OS}^2 \) is $-$0.41\% in our out-of-sample, indicating that the combination of economic variables fails to beat the benchmark of historical average. Additionally, when comparing with the first four principal components of economic variables discussed in Section \ref{Economic Variable}, we find an unreported \( R_{OS}^2 \) of $-$8.22\%, further illustrating the distinct predictive power of \( NR^G \).

\begin{figure}[H] 
\captionsetup{singlelinecheck=false,justification=raggedright,labelfont=bf}
\caption{\textbf{Differences in Cumulative Squared Forecast Errors}}\vspace{-0.2cm}
{\small This figure plots the difference between the cumulative squared forecast error (CSFE) generated by the historical average benchmark forecast and the CSFE derived from forecasts based on good news ratio, $NR^{G}$, which is the monthly proportion of good news identified by ChatGPT-3.5. It answers whether the input news means \textit{GOING UP} for stock market. As a comparison, the figure also depicts the difference in CSFE for the mean combination of forecasts based on the 14 economic variables of \citet*{welch2008comprehensive}. The out-of-sample period spans from January 2006 to December 2022. Grey shadow bars denote NBER recessions. }\label{Fig: CSFE}

\begin{center}
\includegraphics[viewport=1cm 1cm 20cm 23cm]{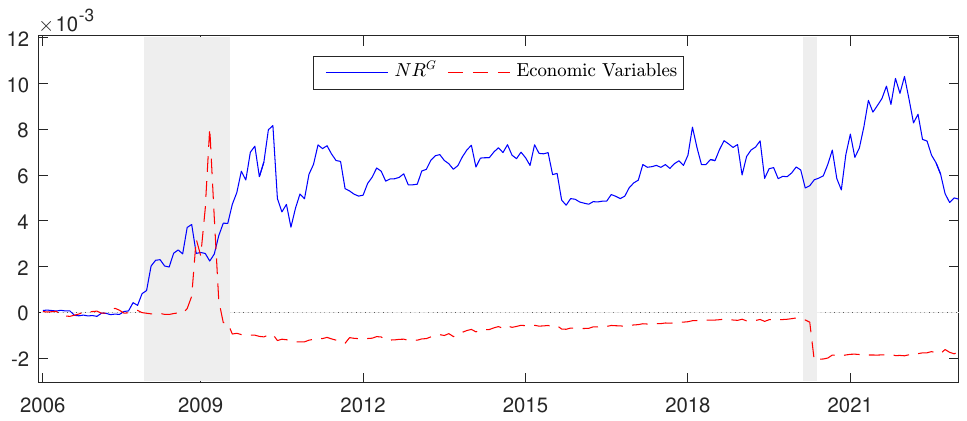}
\end{center}

\end{figure}

\vspace{0.15in}
\centerline{[Insert Figure \ref{Fig: CSFE} about here]}
\vspace{0.05in}

In light of the pronounced out-of-sample predictability of \( NR^G \), an intriguing line of inquiry pertains to whether this predictability is consistent across the entire sample or confined to specific periods. To investigate this, we adopt the approach suggested by \citet*{welch2008comprehensive}, focusing on the temporal evolution of the predictive ability. This involves plotting a time series that represents the difference between the cumulative squared forecast error (CSFE) generated by the historical average benchmark forecast and the CSFE derived from forecasts based on \( NR^G \). A trend characterized by a positive slope in this time-series differential would indicate that \( NR^G \)-based forecasts consistently outperform the historical average across various time periods. This graphical representation thereby provides a vivid illustration of the dynamic performance of \( NR^G \) as a predictor, over time, further enriching our understanding of its reliability and robustness.

Figure \ref{Fig: CSFE} illustrates the difference in CSFE of forecasts based on \( NR^G \) and forecasts of the mean combination of economic variables. Notably, the curve representing \( NR^G \) exhibits a marked upsurge during the periods 2008--2010 and 2021--2022, interspersed with fluctuations across other periods, barring the final year. The overall positive trajectory of the curve signifies a stable predictability of \( NR^G \) over time. The recent downward trend may reflect the increasing availability and influence of LLMs, akin to the publication effect posited by \citet*{mclean2016does}. In contrast, the curve associated with the mean combination of economic variables demonstrates a generally negative slope, punctuated only by a transient increase during the 2008 financial crisis. Collectively, Figure \ref{Fig: CSFE} underscores the enduring predictive capacity of \( NR^G \) across different time frames, thereby affirming its utility in complementing the predictive power inherent in macroeconomic variables.

\subsection{\it \textbf{Economic Value}}
\setlength{\parindent}{2em}
\indent
In this subsection, we delve into the question of whether the out-of-sample forecasting abilities of \( NR^B \) and \( NR^G \), as generated by ChatGPT-3.5, can confer tangible economic benefits to investors. This analysis is particularly pertinent for those contemplating the integration of such forecasting information into their investment strategies, as opposed to disregarding it. We approach this inquiry from the perspective of asset allocation, aiming to quantify the potential economic gains that might accrue from leveraging the predictive insights offered by \( NR^B \) and \( NR^G \).

In alignment with the methodologies advocated by \citet*{kandel1996predictability}, \citet*{campbell2008predicting}, and \citet*{ferreira2011forecasting}, we consider a mean-variance investor who utilizes return forecasts to make his asset allocation decisions between risky stocks and risk-free bills. Portfolio rebalancing is conducted at the end of each month, with the equity weights in the portfolio determined as per the following equation:
\begin{equation}\label{weight}
w_{t}=\frac{1}{\gamma}\,\frac{\widehat{R}_{t+1}}{{\widehat{\sigma}}^{2}_{t+1}} \ ,
\end{equation}
where \( \gamma \) signifies the investor's degree of risk aversion, \( \widehat{R}_{t+1} \) represents the out-of-sample forecast of excess stock returns, and \( {\widehat{\sigma}}^{2}_{t+1} \) is the variance forecast. Consistent with \citet*{campbell2008predicting}, we presume that investors estimate future stock return variances using a 5-year moving window of past returns. Furthermore, the weight \( w_t \) is bounded between 0 and 1.5 to preclude short selling and limit the maximum leverage to 50\%.

The investment strategy entails allocating \(1 - w_{t}\) of the portfolio to risk-free bills. Consequently, the realized portfolio return at time \( t+1 \), denoted as \( R^p_{t+1} \), is expressed by the following equation:
\begin{equation}
R^p_{t+1} = w_{t} \, {R}_{t+1} + R^f_{t+1} \ ,
\end{equation}
where \( R_{t+1} \) represents the excess return of the market portfolio, and \( R^f_{t+1} \) is the risk-free return. The certainty equivalent return (CER) of the portfolio is computed as:
\begin{equation}\label{CER}
CER_p = \widehat{\mu}_{p} - 0.5 \, \gamma \, \widehat{\sigma}^{2}_{p} \ ,
\end{equation}
where \( \widehat{\mu}_{p} \) and \( \widehat{\sigma}^{2}_{p} \) are the sample mean and variance of the investor's portfolio over the forecast evaluation period, respectively. The CER can be interpreted as the risk-free return that an investor would be willing to accept in lieu of holding a risky portfolio. The CER gain is thus the difference between the CER of an investor utilizing the predictive regression forecast of monthly returns as given by Equation (\ref{Pred_OS}) and that of an investor relying on a historical average forecast. This difference, when multiplied by 12, represents the annual portfolio management fee that an investor might be prepared to pay for access to predictive regression forecasts. Additionally, we calculate the annualized Sharpe ratios of \( R^p_{t} \) to further evaluate investment performance. This approach allows us to directly measure the economic value derived from return predictability.

\vspace{0.15in}
\centerline{[Insert Table \ref{Tab: Asset Allocation} about here]}
\vspace{0.05in}

Table \ref{Tab: Asset Allocation} shows the asset allocation results for the out-of-sample period spanning January 2006 to December 2022. The findings reveal that \( NR^G \) achieves significant CER gains, e.g. 4.92\% when risk aversion is three, suggesting that investors might be inclined to pay an annual fee of up to 492 basis points (bps) for access to the predictive regression forecasts based on \( NR^G \). The investment portfolio formulated on the basis of \( NR^G \) reports annualized Sharpe ratios from 0.51 to 0.53, which are substantially higher than the market portfolio's Sharpe ratio of 0.30. This profitability remains considerable even after deducting a proportional transaction cost of 50 basis points, resulting in net-of-transaction-cost CER gains for \( NR^G \), ranging from 1.51\% to 4.73\%. In contrast, \( NR^B \) yields negative CER gains and lower Sharpe ratios, reflecting its relatively weaker forecasting power compared to \( NR^G \).

To summarize, our comprehensive analysis underscores the strong forecasting power of the good news ratio, as identified by ChatGPT, for monthly out-of-sample market returns. This predictive ability translates into significant investment profits within the context of asset allocation, thereby indicating considerable economic value for mean-variance investors. Such results suggest the critical importance of the information extracted by GPT models, especially when viewed from the perspective of asset allocation. The implications of these findings are substantial, revealing the potential utility of incorporating ChatGPT-derived insights into investment strategies.

\subsection{\it \textbf{Look-ahead Bias}}
\setlength{\parindent}{2em}
\indent

One primary concern regarding ChatGPT's outperformance over human analysis, the Bag of Words model, and BERT relates to potential look-ahead bias. Specifically, the training methodology for GPT models might incorporate information not available to humans, the Bag of Words approach, or BERT since the GPT-3.5 is trained based on the text corpus prior to Sep, 2021. To address this issue, we introduce two additional tests.

First, we assess the comparative performance of ChatGPT-3.5 and BERT in predicting monthly out-of-sample returns. Google released BERT in 2018, utilizing only data available up to that year. Our findings indicate that BERT lacks the predictive strength exhibited by ChatGPT in-sample. If ChatGPT-3.5's superior performance persists from 2018 to 2022 and does not significantly differ from the prior periods, it likely stems from the intrinsic capabilities of the GPT model rather than from information leakage. Should look-ahead bias be a significant factor, we would anticipate ChatGPT's outperformance to be confined to the 2018-2021 window. Both models had access to identical information before 2018, and neither should exhibit differential performance after September 2021, as both were restricted from information beyond that point.

\begin{figure}[H] 
\captionsetup{singlelinecheck=false,justification=raggedright,labelfont=bf}
\caption{\textbf{Out-of-sample Comparison with Bert}}\vspace{-0.2cm}
{\small This figure plots the difference between the cumulative squared forecast error (CSFE) generated by the benchmark forecast of Bert and the CSFE derived from forecasts based on good news ratio, $NR^{G}$, which is the monthly proportion of good news identified by ChatGPT-3.5. It answers whether the input news means \textit{GOING UP} for stock market. The out-of-sample period spans from January 2006 to December 2022. Grey shadow bars denote NBER recessions. }\label{Fig: CSFE_bert}

\begin{center}
\includegraphics[viewport=1cm 1cm 20cm 23cm]{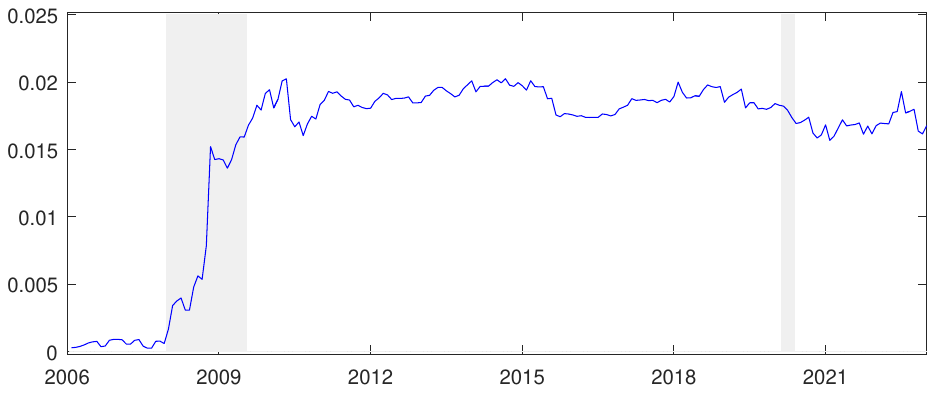}
\end{center}

\end{figure}

Figure \ref{Fig: CSFE_bert} presents the disparity in cumulative sum of forecast errors (CSFE) between BERT's benchmark forecasts and those derived from ChatGPT-3.5, based on the good news ratio $NR^G$, for the out-of-sample period from January 2006 to December 2022. A marked increase during the 2008 financial crisis highlights ChatGPT-3.5's robust predictive ability in turbulent times, discussed further in Section \ref{sec: eco_explanation}. Post-crisis, the curve's flat trajectory suggests ChatGPT's sustained and stable outperformance is due to its superior textual analysis capabilities, not because its training incorporated forward-looking information.

\vspace{0.15in}
\centerline{[Insert Figure \ref{Fig: CSFE_bert} about here]}
\vspace{0.05in}

Secondly, we investigate the return predictability of the news ratio identified by ChatGPT-3.5 from October 2021 through December 2023. According to OpenAI, ChatGPT-3.5 was trained on data available only up to September 1, 2021, implying that any information subsequent to October 2021 was beyond its knowledge base. If our initial findings were predominantly influenced by look-ahead information, the good news ratio's predictive capability would diminish during this timeframe. To test this hypothesis, we constructed weekly good and bad news ratios based on the preceding four weeks and assessed their ability to predict weekly returns. Here, we use weekly rather than monthly news ratio to argument observations due to very limited data after 2021. Utilizing the first 35 weeks for training and the subsequent period for evaluation, Figure IA 1 contrasts the cumulative sum of forecast errors (CSFE) from the historical average benchmark against the CSFEs informed by good (\( NR^G \)) and bad (\( NR^B \)) news ratios. The \( NR^G \) curve's positive trajectory underscores a robust and consistent out-of-sample forecast accuracy. In contrast, the \( NR^B \) curve remains flat, indicating negligible predictive strength. These results, spanning October 2021 to December 2023, affirm the superior predictive performance of the good news ratio over the bad news ratio for weekly market returns, aligning with our prior monthly analyses. Therefore, we conclude that the observed return predictability is not attributable to any look-ahead information within ChatGPT-3.5.

Last but not least, the primary objective of training the GPT model is to enhance semantic representation and improve the predictability of subsequent words or sentences, not to forecast stock market returns. Should outperformance stem from integrating additional data, one would expect GPT-4 to surpass GPT-3.5, as GPT-4 incorporates the most recent information available up to April 2023. Contrarily, our findings indicate that GPT-3.5 delivers similar predictability to GPT-4 in predicting aggregate stock market movements, as demonstrated in Table \ref{Tab: GPT4}.

\section{\textbf{Economic Interpretations}}\label{sec: eco_explanation}
\setlength{\parindent}{2em}
\indent
The findings from our study reveal a distinct pattern in market reactions: a gradual response to good news and a more immediate reaction to bad news, as identified by GPT models. This section explores the potential economic explanations underlying this observed evidence.

\subsection{\it \textbf{Links to Macroeconomic Conditions}}
\setlength{\parindent}{2em}
\indent
Our analysis initially concentrates on deciphering the information content extracted by ChatGPT-3.5. Drawing from the foundational concepts of the Intertemporal Capital Asset Pricing Model (ICAPM) as proposed by \citet{merton1973intertemporal}, we recognize that the excess returns of the market are fundamentally interwoven with macroeconomic conditions, which in turn influence the stochastic nature of the investment opportunity set. In a recent advancement, \citet*{bybee2023narrative} have undertaken an innovative approach to estimate state variables that forecast shifts in future investment opportunities. This estimation is based on narrative elements derived from news text, employing a narrative factor pricing model. Building upon this perspective, we posit that the information extracted by ChatGPT-3.5 from various news narratives is potentially indicative of underlying macroeconomic fundamentals. This hypothesis aligns with the evolving understanding of how modern natural language processing tools, such as ChatGPT-3.5, can offer insightful interpretations of economic indicators and trends from vast textual datasets.

\vspace{0.15in}
\centerline{[Insert Table \ref{Tab: Links to Macro} about here]}
\vspace{0.05in}

To explore this conjecture, we employ several macroeconomic condition proxies, including the Industrial Production Growth (IPG), the CBOE Volatility Index (VIX), the Chicago Fed National Activity Index (CFNAI), the Aruoba-Diebold-Scotti Business Conditions Index (ADSI), the Kansas City Financial Stress Index (KCFSI), Total Non-farm Payroll Growth (Payroll Growth), Smoothed Recession Probability, and Real GDP Growth (GDPG). We regress these proxies on the news ratios as follows:
\begin{equation}\label{economic}
Y_{t+1} = \alpha + \beta\, NR^{K}_t + \epsilon_{t+1} \ , \qquad K = B\ \mbox{or}\ G \ ,
\end{equation}
where \( NR^{K} \) represents the news ratios, either \( NR^{G} \) or \( NR^{B} \). As evidenced in Table \ref{Tab: Links to Macro}, the regression slopes on \( NR^{B} \) are significantly positive for VIX, KCFSI, and SRP, and negative for IPG, CFNAI, and GDPG. This suggests that higher ratio of bad news correlates with heightened market volatility, financial stress, recession probability, and lower industrial production and real GDP growth, indicating powerful ability of ChatGPT to capture the economic downturns. Similarly, \( NR^{G} \) effectively forecasts future macroeconomic conditions, with all regression coefficients being statistically significant. Positive news associates with future high industrial production, real GDP growth, enhanced economic activity, favorable business conditions, employment growth, but lower market volatility and recession probability.

In contrast, when assessing the predictive capabilities of word lists and BERT models (as shown in Table IA 6 of the Internet Appendix), we observe that these methods, although capable of forecasting some macroeconomic variables, exhibit limited economic magnitude and lower \( t \)-statistics. This observation provides further insight into their relatively restricted predictive power in stock market contexts, as discussed in Table \ref{Tab: comparison}.

\vspace{0.15in}
\centerline{[Insert Table \ref{Tab: Links to Sentiment} about here]}
\vspace{0.05in}

DeepSeek has a similar story. As Panel A of Table \ref{Tab: Links to Sentiment} shows, we find that the bad news ratio ($NR^B$) is significantly correlated with most macroeconomic variables, consistent with the results for ChatGPT as shown in Table \ref{Tab: Links to Macro}. In contrast, the good news ratio has limited predictive power. The results are consistent with the low predictive power of DeepSeek on the stock market.

We further compare ChatGPT and DeepSeek in their abilities to discern textual information. We regress the monthly changes in investor sentiment, as proposed by \citet{baker2006investor}, on the news ratios identified by ChatGPT or DeepSeek. Panel B of Table \ref{Tab: Links to Sentiment} shows that, while the bad news identified by both models is significantly correlated with sentiment changes, only DeepSeek's good news significantly affects sentiment. Thus, our findings suggest that DeepSeek might be better at mimicking investor behavior to identify the sentiment component of news, while ChatGPT might be more professional, with its information being correlated with fundamentals.

In summary, this section analyzes the link between large language models (LLMs) and future macroeconomic conditions. A higher ratio of good news identified by ChatGPT is indicative of an improving economic state, whereas a higher ratio of bad news suggests deteriorating future economic conditions. This outcome evidences the capability of ChatGPT in capturing pertinent information about the macroeconomic state from news text. These findings highlight the potential of advanced language models like ChatGPT in offering insightful macroeconomic forecasts based on their analysis of textual data. In contrast, DeepSeek is likely to capture the sentiment component of macroeconomic news, suggesting its ability to mimic investor behavior.


\subsection{\it \textbf{Asymmetric Reactions to News}}
\setlength{\parindent}{2em}
\indent
Our findings indicate a predominant focus of ChatGPT-identified information on macroeconomic fundamentals, eliciting asymmetric market reactions. Specifically, market prices integrate positive news more sluggishly compared to the efficient assimilation of negative news. This subsection delves into the mechanisms underpinning this asymmetry, aiming to elucidate the complex interplay between news content, investor perceptions, and market dynamics.

According to the theoretical framework posited by \citet*{epstein2008ambiguity}, investors perceive information of uncertain quality as ambiguous, leading to deviations from standard Bayesian belief updates. In this context, negative information is often considered highly informative, while positive news is deemed less precise. Consequently, ambiguity-averse investors are prone to discount positive news but place significant weight on negative news. Should this theory hold, a stronger impact of negative over positive news on investor expectations would be observable.

\vspace{0.15in}
\centerline{[Insert Table \ref{Tab: Macro Expectations} about here]}
\vspace{0.05in}

To investigate this phenomenon, we analyze forecasts from the Survey of Professional Forecasters (SPF) concerning key economic fundamentals over subsequent quarters. Our analysis encompasses equal-weighted quarterly forecasts for six pivotal economic indicators: real GDP growth, industrial production growth, unemployment rate, non-farm payroll growth, T-bill yield, and inflation rate.\footnote{For comparability, these six economic variables are standardized before being equal-weighted.} To mitigate look-ahead bias,\footnote{Macroeconomic forecasts are typically released mid-quarter, whereas ChatGPT processes news throughout the quarter.} we incorporate lagged news ratios into our model:
\begin{equation}\label{expectation}
E_{t} = \alpha + \beta\, NR^{K}_{t-1} + \psi\, E_{t-1} + \epsilon_{t} \ , \quad K = B\ \text{or}\ G \ ,
\end{equation}
where \(E_{t}\) denotes the equal-weighted macroeconomic forecasts at time \(t\), and \(NR^{K}\) represents the quarterly news ratio, either for good (\(NR^{G}\)) or bad news (\(NR^{B}\)). The regression outcomes, presented in Table \ref{Tab: Macro Expectations}, underscore a significant negative correlation for the bad news ratio (\(NR^{B}\)) with current economic expectations, save for the forecasts extending four quarters ahead. This finding exemplifies the asymmetric influence of news type on investor expectations, affirming the hypothesis set forth by \citet*{epstein2008ambiguity}.

In essence, our analysis corroborates the heightened sensitivity of investors to negative news, aligning with the theoretical insights of \citet*{epstein2008ambiguity}. This suggests that compared to human investors, ChatGPT exhibits a superior capacity to process positive news, culminating in a lagged market response to such news.

\subsection{\it \textbf{Additional Results for Good News Ratio}}
\setlength{\parindent}{2em}
\indent
Our evidence has shown that the good news ratio ($NR^G$) can predict the stock market positively because investors can not fully capture the textual information. If this interpretation holds, we would expect to observe stronger return predictability of $NR^G$ during periods when the news is more challenging for human investors to comprehend. We consider three scenarios: (1) Human investors may not fully capture the information of good news in bad times; (2) Human investors may not fully capture the information if it is ambiguous with respect to its implication; (3) Human investors may not fully capture the information if it is novel.

\subsubsection{\it\large \textbf{Interaction with Economic States}}
\setlength{\parindent}{2em}
\indent
The business cycle's current state significantly influences market reactions to media news. Particularly during recessionary periods, investor uncertainty regarding future economic growth intensifies, potentially leading to incomplete assimilation of positive news. \citet{veronesi1999stock} underscores this phenomenon, noting a pronounced underreaction of stock prices to favorable news amidst economic downturns. Given this context, ChatGPT's ability to identify positive news is hypothesized to yield stronger market predictions, attributing to the lesser degree of information integration by investors during such times.

\vspace{0.15in}
\centerline{[Insert Table \ref{Tab: Econ Activity} about here]}
\vspace{0.05in}

To examine the relationship between economic activity and market response to news, we utilize the Chicago Fed National Activity Index (CFNAI) as a proxy for economic conditions, assessing overall economic activity and related inflationary pressures. An indicator variable \( I_{High} \) is constructed, assigned a value of one if the current CFNAI exceeds the past five-year sample mean, and zero otherwise. \( I_{Low} \) is defined as \( 1 - I_{High} \). The predictive regression model is formulated as follows:
\begin{equation}\label{activity}
R_{t+h} = \alpha + \beta_1 \, I_{High} \times NR^G_t + \beta_2 \, I_{Low} \times NR^G_t + \beta_3 \, I_{High} + \epsilon_{t+h} \ ,
\end{equation}
where \( R_{t+h} \) represents the average excess return of the market portfolio from \( t+1 \) to \( t+h \), for \( h = 1, 3, 6, 9, \) and \( 12 \) months. The variable \( NR^G \) denotes the good news ratio, as identified by GPT-3.5. We primarily focus on the coefficients of \( I_{High} \times NR^G_t \) and \( I_{Low} \times NR^G_t \). The statistical significance of these coefficients will indicate the predominant periods (high or low economic activity) during which the return predictability of \( NR^G_t \) is more pronounced. As depicted in Table \ref{Tab: Econ Activity}, our results reveal that while the coefficient for \( I_{High} \times NR^G_t \) remains insignificant across various horizons, the coefficient for \( I_{Low} \times NR^G_t \) ranges from 0.71\% to 0.97\%, achieving statistical significance at the 10\% level or better. This outcome corroborates the findings of \citet{veronesi1999stock} and \citet*{garcia2013sentiment}, indicating that the return predictability of \( NR^G_t \) is markedly more substantial during periods of economic downturns, aligning with their observations regarding the heightened predictive power of news content during recessions. In an unported table, we use the NBER-dated recession to construct the economic condition dummy and find similar results that the predictability of $NR^G$ is stronger in recessions.

\subsubsection{\it\large \textbf{Interaction with Economic Policy Uncertainty}}
\setlength{\parindent}{2em}
\indent
Ambiguity in the information processed by human investors often results in challenges in comprehension. Such ambiguity, referred to as information uncertainty by \citet{zhang2006information}, is hypothesized to enhance the predictive power of ChatGPT-identified information. To quantify this uncertainty, we utilize the Economic Policy Uncertainty (EPU) index, introduced by \citet*{baker2016measuring}, as a measure of information uncertainty. This study aims to assess the predictive capabilities of the good news ratio (\( NR^G \)) under varying levels of EPU.

\vspace{0.15in}
\centerline{[Insert Table \ref{Tab: EPU} about here]}
\vspace{0.05in}

We construct an indicator variable, \( U_{High} \), which takes the value of one if the current EPU exceeds the past five-year sample mean, and zero otherwise. The counterpart variable, \( U_{Low} \), is defined as \( 1 - U_{High} \). The ensuing regression model is specified to evaluate the impact of these EPU-related indicators on the return predictability of the good news ratio (\( NR^G \)):
\begin{equation}\label{uncertainty}
R_{t+h} = \alpha + \beta_1 \, U_{High} \times NR^G_t + \beta_2 \, U_{Low} \times NR^G_t + \beta_3 \, U_{High} + \epsilon_{t+h} \ .
\end{equation}
As depicted in Table \ref{Tab: EPU}, the estimation results of this regression model (\ref{uncertainty}) highlight the influence of EPU on the return predictability of \( NR^G \). The coefficient for the interaction term \( U_{High} \times NR^G_t \) is statistically significant at the 5\% level or better, with values ranging from 0.78\% to 0.89\%. Given that all predictors are standardized with zero mean and unit variance, the economic magnitude of these coefficients implies an increase in returns by 0.78\% to 0.89\% following a one-standard-deviation increase in \( NR^G \) during periods of high EPU. This finding is substantially more pronounced than the results observed during periods of low EPU, which peak at a maximum of 0.29\%.

\subsubsection{\it\large \textbf{Interaction with News Similarity}}
\setlength{\parindent}{2em}
\indent
The novelty of news content is crucial in shaping investor understanding of public information. When confronted with new information, investors may not fully process its implications \citep*{chan1996momentum}. Consequently, this suggests that ChatGPT's ability to identify and interpret novel news may confer enhanced predictive power for stock market forecasting. The distinctiveness of newly released information, therefore, is likely to be a significant determinant of ChatGPT's forecasting efficacy. In a related vein, \citet*{tetlock2011all} approached the concept of news novelty by assessing the similarity of current news stories to prior ones regarding the same entity. In line with \citet*{tetlock2011all}'s methodology, we adopt a similar approach for evaluating the novelty of economic-relevant news, as delineated in Section \ref{similarity}. Here, "economic-relevant news" encompasses stories featuring economic-related keywords, which are detailed in \ref{appendix_b}. This analysis aims to understand how the freshness or staleness of economic news impacts investor behavior and market responses.

\vspace{0.15in}
\centerline{[Insert \ref{Tab: Similarity} about here]}
\vspace{0.05in}

The regression model constructed to assess the influence of news novelty on market returns is formulated as follows:
\begin{equation}\label{news_similarity}
R_{t+h} = \alpha + \beta_1 \, S_{High} \times NR^G_t + \beta_2 \, S_{Low} \times NR^G_t + \beta_3 \, S_{High} + \epsilon_{t+h} \ ,
\end{equation}
where \( S_{High} \) represents an indicator variable that takes the value of one when the current news similarity exceeds the past five-year sample mean, and zero otherwise. The counterpart variable \( S_{Low} \) is defined as \( 1 - S_{High} \). As detailed in Table \ref{Tab: Similarity}, the estimation results for regression (\ref{news_similarity}) reveal notable findings. The coefficient for the interaction term \( S_{High} \times NR^G_t \) does not demonstrate statistical significance across different time horizons. Conversely, the coefficient for \( S_{Low} \times NR^G_t \) varies between 0.67\% and 0.84\%, with \( t \)-statistics ranging from 2.19 to 3.04. This pattern suggests that the predictability of \( NR^G_t \) is more evident when the economic news is particularly novel compared to prior reports.

In summary, the evidence gathered from our analysis in this section indicates a notable delay in market response to good news as identified by the ChatGPT model. This evidence is particularly pronounced during economically challenging times and during periods characterized by high EPU, indicating the influence of broader economic conditions on market responses to public information. Moreover, the predictability of ChatGPT is stronger for the unique novelty of news.

\section{\textbf{Conclusions}}\label{sec: conclusion}
\setlength{\parindent}{2em}
\indent
In this paper, we employ ChatGPT to extract good and bad news regarding the stock market from both the news headlines and alerts on the \emph{Wall Street Journal} from 1996 to 2022, and compare it with DeepSeek and other LLMs. Our findings reveal a notable association between a high percentage of identified good news and subsequent market returns at a the monthly frequency. This return predictability extends to the next six months and is robust to various prompt setups. Moreover, we find that the text information identified by ChatGPT is more likely to be related to macroeconomic conditions. We also find that the "large" LLMs have the superior ability over the small LLMs, like BERT, and over the traditional text analysis method that usually assigns a context-independent representation to each word or sentence. We also find that DeepSeek underperforms ChatGPT, due to perhaps the former is trained primarily on Chinese and English datasets while the latter is trained on multiple languages focusing on English\footnote{An on going project of the authors is to apply our methods here to Chinese textual data to predict the Chinese stock market, to see which of the two outperforms.}.

Our results show that the ChatGPT has emergent abilities in identifying good news that go beyond the comprehension of human investors. This "over-performance" becomes statistically and economically more significant during economic downturns, rising economic policy uncertainty, and flourishing news novelty. Conversely, our analysis reveals that human investors do exhibit a relatively efficient capacity to assess, interpret and assimilate bad news. This finding in the efficacy of information digestion between positive and negative news is consistent with the theoretical model of \citet{epstein2008ambiguity}, and underscores the nuanced nature of investor responses to varying news types that may have wide implications.

In short, our study establishes ChatGPT's market prediction capabilities, shedding light on the contrasting abilities of LLMs and human investors in processing text news, particularly under stressful market conditions. These insights contribute to our improved understanding of the interplay between LLMs and human interpretation in the realm of financial market information. Future research is called for to apply LLMs to other financial markets, such as bonds, currencies and commodities, to learn how information processing of ChatGPT and investors differs and what impact they have on asset prices and investment performance.

\clearpage \pagebreak

\appendix
\renewcommand{\thesection}{\textbf{Appendix \Alph{section}}}

\section{\textbf{Description for Economic Variables}}\label{appendix_a}
\setcounter{equation}{0}
\renewcommand\theequation{B.\arabic{equation}}
\indent
The 14 economic variables of \citet*{welch2008comprehensive} are defined as,
\renewcommand{\labelitemi}{\textbullet}
\begin{itemize}
\item
Dividend-price ratio (log), DP: log of a twelve-month moving sum of dividends paid on the S\&P 500 index minus the log of stock prices (S\&P 500 index).

\item
Dividend yield (log), DY: log of a twelve-month moving sum of dividends minus the log of lagged stock prices.

\item
Earnings-price ratio (log), EP: log of a twelve-month moving sum of earnings on the S\&P 500 index minus the log of stock prices.

\item
Dividend-payout ratio (log), DE: log of a twelve-month moving sum of dividends minus the log of a twelve-month moving sum of earnings.

\item
Stock return variance, SVAR: sum of squared daily returns on the S\&P 500 index.

\item
Book-to-market ratio, BM: ratio of book value to market value for the Dow Jones Industrial Average.\footnote{We compute the logarithm of the book-to-market ratio in the empirical analysis.}

\item
Net equity expansion, NTIS: ratio of a twelve-month moving sum of net equity issues by NYSE-listed stocks to the total end-of-year market capitalization of NYSE stocks.

\item
Treasury bill rate, TBL: interest rate on a three-month Treasury bill (secondary market).

\item
Long-term yield, LTY: long-term government bond yield.

\item
Long-term return, LTR: return on long-term government bonds.

\item
Term spread, TMS: long-term yield minus the Treasury bill rate.

\item
Default yield spread, DFY: difference between BAA- and AAA-rated corporate bond yields.

\item
Default return spread, DFR: long-term corporate bond return minus the long-term government bond return.

\item
Inflation, INFL: calculated from the CPI for all urban consumers; we use lagged two-month inflation in regression to account for the delay in CPI releases.

\end{itemize}

\section{\textbf{Lists of Economic Keywords}}\label{appendix_b}
\setcounter{equation}{0}
\renewcommand\theequation{B.\arabic{equation}}
\indent
The economic-relevant news is the news that includes the following keywords:\\[1.5ex]
\textit{Dow Jones, stock exchange, stock prices, stock market, Nasdaq market, Nasdaq stock, security exchange, security price, security market, interest rate, debt market, security, market, economy, fed, bank, finance, monetary}

\clearpage \pagebreak

\newpage

\bibliographystyle{chicago}
\bibliography{reference}


\clearpage\pagebreak

\clearpage\pagebreak


\newpage
\begin{landscape}
\begin{table} [!htbp]
\captionsetup{singlelinecheck=false,justification=raggedright,labelfont=bf}
\caption{\textbf{Summary Statistics of \emph{Wall Street Journal} News }} \vspace{-0.2cm}
{\small This table reports the mean, standard deviation (Std. Dev.), skewness (Skew.), median, minimum (min.) and maximum (Max.) of the total monthly news, monthly bad news, monthly neutral news, and monthly good news. The good, neutral, or bad news is identified by ChatGPT-3.5 in Panel A and by DeepSeek-R1 in Panel B. It answers whether the input news means \textit{GOING UP}, \textit{GOING DOWN}, or \textit{UNKNOWN} for stock market. In the last column, we present the number of news in our sample from January 1996 to December 2022. }\label{Tab: summary}

\renewcommand{\arraystretch}{1.6}
\tabcolsep 22pt
\small
\centering
\begin{tabular}{l ccccccc} \\[-15pt]\toprule\\[-20pt]

& Mean  & Std. Dev. & \hspace{0.2cm} Skew. & \hspace{0.1cm}Median & Min.  & Max. & No. of News \\[-3pt]\midrule\\[-20pt]

Total News	&	260.91 	&	116.12 	&	$-$0.22 	&	288	&	47	&	577	&	84535	\\[-3pt] \hline\\[-15pt]

\multicolumn{7}{l}{\underline{Panel A: Results for ChatGPT}}\\

\emph{Bad} News	&	32.76 	&	23.93 	&	1.15 	&	32	&	0	&	142	&	10613	\\
\emph{Neutral} News	&	181.75 	&	74.34 	&	$-$0.07 	&	190	&	43	&	422	&	58888	\\
\emph{Good} News	&	46.40 	&	28.25 	&	0.10 	&	47	&	0	&	121	&	15034	\\[-3pt] \hline\\[-15pt]

\multicolumn{7}{l}{\underline{Panel B: Results for DeepSeek}}\\

\emph{Bad} News	&	37.51	&	24.43	&	0.93	&	38	&	1	&	137	&	12153	\\
\emph{Neutral} News	&	180.46	&	77.40	&	$-$0.16	&	194	&	35	&	414	&	58469	\\
\emph{Good} News	&	42.94	&	24.75	&	0.10	&	43	&	1	&	111	&	13913	\\[-3pt]
\bottomrule

\end{tabular}
\end{table}
\end{landscape}

\clearpage \pagebreak


\normalsize\newpage
\begin{table} [!htbp]
\captionsetup{singlelinecheck=false,justification=raggedright,labelfont=bf}
\caption{\textbf{Results of Baseline Regressions }} \vspace{-0.2cm}
{\small This table reports estimation results of the following regression,\vspace{-0.1cm}
$$ R_{t+h} = \alpha + \beta\, NR_{t}^{K} + \epsilon_{t+h} \ , \qquad K=B\ \mbox{or}\ G \ ,  \vspace{-0.2cm} $$
where $R_{t}$ denotes the current market excess return on the S\&P 500 index at time $t$ for "$h=0$". This setting aligns with a contemporaneous regression framework. For scenarios where $h>0$, $R_{t+h}$ represents the average excess returns of the market portfolio from $t+1$ to $t+h$ (with $h$ being 1, 3, 6, 9, and 12 months), transitioning the equation into a predictive regression. $NR^{K}$ represents news ratios. Specifically, $NR^{B}$ is the monthly proportion of bad news and $NR^{G}$ represents the proportion of good news. The good or bad news is identified by ChatGPT-3.5 in Panel A and by DeepSeek-R1 in Panel B. It answers whether the input news means \textit{GOING UP} or \textit{GOING DOWN} for the stock market. Reported are the regression slopes and $R^2$s in percentage form. Also reported are the \citet{hodrick1992dividend} $t$-statistics. All the forecast variables are standardized to have a zero mean and unit variance. *, **, and *** indicate significance at the 10\%, 5\%, and 1\% levels, respectively. The sample period is from January 1996 to December 2022.  } \label{Tab: In-sample}

\renewcommand{\arraystretch}{1.6}
\tabcolsep 14pt
\small
\centering

\begin{tabular}{l r@{.}lr@{.}lr@{.}l c r@{.}lr@{.}lr@{.}l}\\[-15pt] \toprule\\[-20pt]

& \multicolumn{6}{c}{\emph{Bad} News Ratio ($NR^B$)} && \multicolumn{6}{c}{\emph{Good} News Ratio ($NR^G$)}\\ \cline{2-7}\cline{9-14}\\[-20pt]

& \multicolumn{2}{c}{$\beta$ (\%)}   & \multicolumn{2}{c}{Hodrick-$t$}        & \multicolumn{2}{c}{$R^2$ (\%)} && \multicolumn{2}{c}{$\beta$ (\%)}   & \multicolumn{2}{c}{Hodrick-$t$}        & \multicolumn{2}{c}{$R^2$ (\%)}\\[-3pt] \midrule \\[-20pt]

\multicolumn{14}{l}{\underline{Panel A: Results for ChatGPT}}\\[3pt]

$h=$ 0	&	$-$1&03$^{***}$	&	\hspace{0.1cm}$-$2&70 	&	\hspace{0.1cm}5&17 			&&	1&02$^{***}$	&	\hspace{0.4cm}5&30 	&	\hspace{0.1cm}5&07 	\\	
$h=$ 1	&	0&05 		&	0&14 	&	0&01 		&&	0&53$^{**}$	&	2&22 	&	1&37 	\\	
$h=$ 3	&	0&06 		&	0&18 	&	0&05 		&&	0&56$^{**}$	&	2&34 	&	4&60 	\\	
$h=$ 6	&	0&14 		&	0&47 	&	0&48 		&&	0&51$^{*}$	&	1&84 	&	6&91 	\\	
$h=$ 9	&	0&21 		&	0&76 	&	1&57 		&&	0&44 		&	1&51 	&	7&46 	\\
$h=$ 12	&	0&25 		&	0&94 	&	2&65 		&&	0&42 		&	1&48 	&	8&52 	\\
[-3pt]\midrule\\[-15pt]

\multicolumn{14}{l}{\underline{Panel B: Results for DeepSeek}}\\[3pt]

$h=$ 0	&	$-$1&37$^{***}$ 	&	$-$4&07 	&	9&12 	&&	0&85$^{***}$ 	&	3&64 	&	3&54 	\\
$h=$ 1	&	$-$0&21 	&	$-$0&64 	&	0&22 	&&	0&03 	&	0&11 	&	0&00 	\\
$h=$ 3	&	$-$0&33 	&	$-$0&96 	&	1&60 	&&	0&22 	&	1&24 	&	0&71 	\\
$h=$ 6	&	$-$0&35 	&	$-$1&00 	&	3&03 	&&	0&24 	&	1&28 	&	1&57 	\\
$h=$ 9	&	$-$0&27 	&	$-$0&78 	&	2&51 	&&	0&25 	&	1&22 	&	2&35 	\\
$h=$ 12	&	$-$0&17 	&	$-$0&53 	&	1&23	&&	0&27 	&	1&30 	&	3&70 	\\
[-3pt]
\bottomrule

\end{tabular}
\end{table}

\clearpage \pagebreak

\begin{table} [!htbp]
\captionsetup{singlelinecheck=false,justification=raggedright,labelfont=bf}
\caption{\textbf{Comparisons with Alternative Textual Analysis Methods}}\vspace{-0.2cm}
{\small This table reports estimation results of the following regression,
$$ R_{t+h} = \alpha + \beta\, NR_{t}^{K} + \epsilon_{t+h} \ , \qquad K=B\ \mbox{or}\ G \ ,  \vspace{-0.2cm} $$
where $R_{t}$ denotes the current market excess return on the S\&P 500 index at time $t$ for "$h=0$". This setting aligns with a contemporaneous regression framework. For scenarios where $h>0$, $R_{t+h}$ represents the average excess returns of the market portfolio from $t+1$ to $t+h$ (with $h$ being 1, 3, 6, 9, and 12 months), transitioning the equation into a predictive regression. $NR^{K}$ represents news ratios. Specifically, $NR^{B}$ is the monthly proportion of bad news and $NR^{G}$ represents the proportion of good news. We identify the good or bad news by using the method of word lists proposed by \citet*{loughran2011liability} or using BERT. Reported are the regression slopes and $R^2$s in percentage form. Also reported are the \citet{hodrick1992dividend} $t$-statistics. All the forecast variables are standardized to have a zero mean and unit variance. *, **, and *** indicate significance at the 10\%, 5\%, and 1\% levels, respectively. The sample period is from January 1996 to December 2022.}\label{Tab: comparison}

\renewcommand{\arraystretch}{1.6}
\tabcolsep 14pt
\small
\centering

\begin{tabular}{l r@{.}lr@{.}lr@{.}l c r@{.}lr@{.}lr@{.}l}\\[-15pt] \toprule\\[-20pt]

& \multicolumn{6}{c}{Panel A: Results for $NR^{B}$}  && \multicolumn{6}{c}{Panel B: Results for $NR^{G}$} \\ \cline{2-7}\cline{9-14} \\[-20pt]

& \multicolumn{2}{c}{$\beta$ (\%)}   & \multicolumn{2}{c}{Hodrick-$t$}        & \multicolumn{2}{c}{$R^2$ (\%)} && \multicolumn{2}{c}{$\beta$ (\%)}   & \multicolumn{2}{c}{Hodrick-$t$}        & \multicolumn{2}{c}{$R^2$ (\%)} \\[-3pt] \midrule \\[-20pt]

\multicolumn{14}{l}{\underline{Word Lists}} \\[3pt]

\(h=0\) & $-$0&77\(^{***}\) & \hspace{0.1cm}$-$2&68 & \hspace{0.2cm}2&90 && 0&24 & 1&05 & 0&27 \\
\(h=1\) & $-$0&17 & $-$0&47 & 0&14 && $-$0&23 & \hspace{0.2cm}$-$0&87 & \hspace{0.2cm}0&27 \\
\(h=3\) & $-$0&14 & $-$0&56 & 0&28 && $-$0&14 & $-$0&77 & 0&27 \\
\(h=6\) & 0&05 & 0&27 & 0&06 && $-$0&09 & $-$0&74 & 0&20 \\
\(h=9\) & 0&09 & 0&46 & 0&31 && $-$0&14 & $-$1&38 & 0&70 \\
\(h=12\) & 0&11 & 0&49 & 0&56 && $-$0&11 & $-$1&36 & 0&60
\\[-3pt]\midrule\\[-15pt]

\multicolumn{14}{l}{\underline{Bert}} \\[3pt]

\(h=0\) & $-$0&38 & $-$1&03 & 0&71 && $-$0&55\(^{*}\) & $-$1&87 & 1&49 \\
\(h=1\) & 0&10 & 0&36 & 0&05 && 0&32 & 1&00 & 0&49 \\
\(h=3\) & 0&13 & 0&71 & 0&25 && 0&22 & 0&82 & 0&72 \\
\(h=6\) & 0&18 & 0&86 & 0&81 && 0&23 & 0&82 & 1&38 \\
\(h=9\) & 0&09 & 0&43 & 0&29 && 0&22 & 0&84 & 1&68 \\
\(h=12\) & 0&10 & 0&53 & 0&41 && 0&32 & 1&32 & 4&12 \\[-3pt]
\bottomrule

\end{tabular}
\end{table}

\clearpage \pagebreak

\begin{table} [!htbp]
\captionsetup{singlelinecheck=false,justification=raggedright,labelfont=bf}
\caption{\textbf{Results for Alternative Prompts}}\vspace{-0.2cm}
{\small This table reports estimation results of the following regression,
$$ R_{t+h} = \alpha + \beta\,NR_{t}^K + \epsilon_{t+h}\ , \qquad K=B\ \mbox{or}\ G \ ,  \vspace{-0.2cm} $$
where $R_{t}$ denotes the current market excess return on the S\&P 500 index at time $t$ for "$h=0$". This setting aligns with a contemporaneous regression framework. For scenarios where $h>0$, $R_{t+h}$ represents the average excess returns of the market portfolio from $t+1$ to $t+h$ (with $h$ being 1, 3, 6, 9, and 12 months), transitioning the equation into a predictive regression. $NR^{K}$ represents news ratios. Specifically, $NR^{B}$ is the monthly proportion of bad news and $NR^{G}$ represents the proportion of good news. The good or bad news is identified by ChatGPT-3.5. It answers whether the input news is \textit{OPTIMISTIC} (\textit{PESSIMISTIC}) news for the stock market, or is \textit{POSITIVE} (\textit{NEGATIVE}) news for the stock market. Reported are the regression slopes and \( R^2 \)s in percentage form. Also reported are the \citet{hodrick1992dividend} $t$-statistics. All the forecast variables are standardized to have a zero mean and unit variance. *, **, and *** indicate significance at the 10\%, 5\%, and 1\% levels, respectively. The sample period is from January 1996 to December 2022.} \label{Tab: Prompts}

\renewcommand{\arraystretch}{1.6}
\tabcolsep 14pt
\small
\centering

\begin{tabular}{l r@{.}lr@{.}lr@{.}l c r@{.}lr@{.}lr@{.}l}\\[-15pt] \toprule\\[-20pt]

& \multicolumn{2}{c}{$\beta$ (\%)}   & \multicolumn{2}{c}{Hodrick-$t$}        & \multicolumn{2}{c}{$R^2$ (\%)} && \multicolumn{2}{c}{$\beta$ (\%)}   & \multicolumn{2}{c}{Hodrick-$t$}        & \multicolumn{2}{c}{$R^2$ (\%)} \\[-3pt] \midrule \\[-20pt]

& \multicolumn{6}{l}{\underline{Panel A: \textit{Pessimistic} News}} && \multicolumn{6}{l}{\underline{Panel B: \textit{Optimistic} News}}\\[3pt]

\(h=0\) & $-$0&84\(^{**}\) & \hspace{0.1cm}$-$2&34 & \hspace{0.1cm}3&46 && 1&09\(^{***}\) & \hspace{0.4cm}5&55 & 5&76 \\
\(h=1\) & 0&03 & 0&10 & 0&01 && 0&43\(^{*}\) & 1&92 & 0&88 \\
\(h=3\) & 0&07 & 0&20 & 0&06 && 0&54\(^{**}\) & 2&52 & 4&42 \\
\(h=6\) & 0&11 & 0&36 & 0&29 && 0&52\(^{**}\) & 2&07 & 7&37 \\
\(h=9\) & 0&18 & 0&62 & 1&13 && 0&49\(^{*}\) & 1&80 & 9&30 \\
\(h=12\) & 0&25 & 0&90 & 2&80 && 0&47\(^{*}\) & 1&71 & 10&99
\\[-3pt]\midrule\\[-15pt]

& \multicolumn{6}{l}{\underline{Panel C: \textit{Negative} News}} && \multicolumn{6}{l}{\underline{Panel D: \textit{Positive} News}}\\[3pt]

\(h=0\) & $-$0&49 & $-$1&57 & 1&15 && 0&74\(^{***}\) & 3&31 & 2&65 \\
\(h=1\) & $-$0&04 & $-$0&13 & 0&01 && 0&43\(^{*}\) & 1&84 & 0&88 \\
\(h=3\) & 0&03 & 0&10 & 0&01 && 0&45\(^{**}\) & 1&97 & 3&00 \\
\(h=6\) & 0&10 & 0&39 & 0&27 && 0&51\(^{**}\) & 1&96 & 7&04 \\
\(h=9\) & 0&20 & 0&74 & 1&45 && 0&47 & 1&62 & 8&41 \\
\(h=12\) & 0&28 & 1&00 & 3&72 && 0&45 & 1&50 & 10&00 \\[-3pt]
\bottomrule
\end{tabular}
\end{table}

\clearpage \pagebreak

\begin{table} [!htbp]
\captionsetup{singlelinecheck=false,justification=raggedright,labelfont=bf}
\caption{\textbf{Results for ChatGPT-3.5 Fine--tuning and ChatGPT-4}}\vspace{-0.2cm}
{\small This table reports estimation results of the following regression,
$$ R_{t+h} = \alpha + \beta\,NR_{t}^K + \epsilon_{t+h}\ , \qquad K=B\ \mbox{or}\ G \ ,  \vspace{-0.2cm} $$
where $R_{t}$ denotes the current market excess return on the S\&P 500 index at time $t$ for "$h=0$". This setting aligns with a contemporaneous regression framework. For scenarios where $h>0$, $R_{t+h}$ represents the average excess returns of the market portfolio from $t+1$ to $t+h$ (with $h$ being 1, 3, 6, 9, and 12 months), transitioning the equation into a predictive regression. $NR^{K}$ represents news ratios. Specifically, $NR^{B}$ is the monthly proportion of bad news and $NR^{G}$ represents the proportion of good news. The news is identified by ChatGPT-3.5 fine--tuning and ChatGPT-4, respectively. It answers whether the input news means \textit{GOING UP} or \textit{GOING DOWN} for the stock market. Reported are the regression slopes and \( R^2 \)s in percentage form. Also reported are the \citet{hodrick1992dividend} $t$-statistics. *, **, and *** indicate significance at the 10\%, 5\%, and 1\% levels, respectively. The sample period is from January 1996 to December 2022.}\label{Tab: GPT4}

\renewcommand{\arraystretch}{1.6}
\tabcolsep 14pt
\small
\centering

\begin{tabular}{l r@{.}lr@{.}lr@{.}l c r@{.}lr@{.}lr@{.}l}\\[-15pt] \toprule\\[-20pt]

& \multicolumn{6}{c}{Panel A: Results for $NR^{B}$}  && \multicolumn{6}{c}{Panel B: Results for $NR^{G}$} \\ \cline{2-7}\cline{9-14} \\[-20pt]

& \multicolumn{2}{c}{$\beta$ (\%)}   & \multicolumn{2}{c}{Hodrick-$t$}        & \multicolumn{2}{c}{$R^2$ (\%)} && \multicolumn{2}{c}{$\beta$ (\%)}   & \multicolumn{2}{c}{Hodrick-$t$}        & \multicolumn{2}{c}{$R^2$ (\%)} \\[-3pt] \midrule \\[-20pt]

\multicolumn{14}{l}{\underline{ChatGPT-3.5 Fine--tuning}} \\[3pt]

\(h=0\)	&	$-$1&38\(^{***}\)	&	\hspace{0.1cm}$-$4&69 	&	\hspace{0.1cm}9&20 	&&	0&80\(^{***}\)	&	\hspace{0.4cm}4&04 	&	\hspace{0.2cm}3&11 	\\	
\(h=1\)	&	0&14 		&	0&48 	&	0&09 	&&	0&50\(^{**}\)	&	2&25 	&	1&23 	\\
\(h=3\)	&	0&26 		&	0&82 	&	0&93 	&&	0&43\(^{**}\)	&	2&44 	&	2&71 	\\
\(h=6\)	&	0&21 		&	0&71 	&	1&05 	&&	0&46\(^{**}\)	&	2&57 	&	5&83 	\\
\(h=9\)	&	0&17 		&	0&57 	&	0&92 	&&	0&39\(^{**}\)	&	2&09 	&	5&74 	\\
\(h=12\)	&	0&21 		&	0&81 	&	1&72 	&&	0&34\(^{**}\)	&	2&16 	&	5&75 	\\[-3pt]\midrule\\[-15pt]

\multicolumn{14}{l}{\underline{ChatGPT-4}} \\[3pt]

\(h=0\) & $-$1&08\(^{***}\) & $-$3&68 & 5&62 && 0&90\(^{***}\) & 4&20 & 3&90 \\
\(h=1\) & $-$0&14 & $-$0&40 & 0&10 && 0&37 & 1&56 & 0&66 \\
\(h=3\) & $-$0&09 & $-$0&25 & 0&13 && 0&36\(^{*}\) & 1&83 & 1&90 \\
\(h=6\) & $-$0&07 & $-$0&22 & 0&14 && 0&36\(^{*}\) & 1&83 & 3&53 \\
\(h=9\) & $-$0&01 & $-$0&03 & 0&00 && 0&32 & 1&61 & 3&95 \\
\(h=12\) & 0&09 & 0&32 & 0&38 && 0&28 & 1&60 & 3&81 \\[-3pt]
\bottomrule
\end{tabular}
\end{table}

\clearpage \pagebreak


\begin{table} [!htbp]
\captionsetup{singlelinecheck=false,justification=raggedright,labelfont=bf}
\caption{\textbf{Out-of-sample Forecasting Results  }}\vspace{-0.2cm}
{\small This table reports the out-of-sample $R^2_{OS}$'s, \emph{MSFE-adjusted} statistics, and corresponding $p$-values for predicting the monthly stock market returns based on news ratios ($NR^{B}$ or $NR^{G}$). $NR^{G}$ ($NR^{B}$) is the monthly proportion of good (bad) news identified by ChatGPT-3.5. It answers whether the input news means \textit{GOING UP} or \textit{GOING DOWN} for stock market. Additionally, we report the results for mean combination (MC) of $NR^{B}$ and $NR^{G}$, and the results for iterated mean combination (IMC) and the iterated weighted combination (IWC) proposed by \citet*{lin2018forecasting}. In the last row, we also present the results for mean combination of the forecasts based on 14 economic variables proposed by \citet*{welch2008comprehensive}. The regression slopes are estimated recursively using the data available at the forecast formation time $t$. *, **, and *** indicate significance at the 10\%, 5\%, and 1\% levels, respectively. The out-of-sample evaluation period is from January 2006 to December 2022. }\label{Tab: OOS}

\renewcommand{\arraystretch}{1.6}
\tabcolsep 28pt
\small
\centering
\begin{tabular}{l r@{.}lr@{.}lr@{.}l} \\[-15pt] \toprule\\[-20pt]

&   \multicolumn{2}{c}{$R_{OS}^2$ (\%)}  &    \multicolumn{2}{c}{\emph{MSFE-adjusted} } & \multicolumn{2}{c}{$p$-value }    \\\midrule\\[-15pt]

$NR^{G}$	&	1&17$^{**}$ 	&	2&17 	&	0&01 	\\
$NR^{B}$	&	 $-$2&55 	&	\hspace{0.6cm} $-$1&07 	&	 \hspace{0.4cm}0&86 	\\
MC	&    0&39 	&    1&03     &	  0&15 \\
IMC	   &    1&36$^{**}$   &	1&70    &	0&04 \\
IWC    & 	2&51$^{***}$   &	2&33    &	0&01 \\
Economic Variables & $-$0&41  &	0&07 &	0&47 \\[-3pt]\bottomrule\\

\end{tabular}
\end{table}

\clearpage \pagebreak


\begin{table} [!htbp]
\captionsetup{singlelinecheck=false,justification=raggedright,labelfont=bf}
\caption{\textbf{Economic Values }}\vspace{-0.2cm}
{\small This table presents the CER gains in percentage points and annualized Sharpe ratio for a mean-variance investor who allocates assets between the market portfolio and risk-free bills, with a risk-aversion coefficient ($\gamma$) of one, three, or five. The stock market return forecasts are generated by the regression model based on news ratios ($NR^{B}$ or $NR^{G}$). $NR^{G}$ ($NR^{B}$) is the monthly proportion of good (bad) news identified by ChatGPT-3.5. It answers whether the input news means \textit{GOING UP} or \textit{GOING DOWN} for stock market. A proportional transaction cost (TC) of 50bp is also considered. The out-of-sample evaluation period is from January 2006 to December 2022. }\label{Tab: Asset Allocation}

\renewcommand{\arraystretch}{1.6}
\tabcolsep 26pt
\small
\centering
\begin{tabular}{l r@{.}lr@{.}lr@{.}l} \\[-15pt]\toprule\\[-20pt]

& \multicolumn{2}{c}{CER gain (\%) }  &  \multicolumn{2}{c}{CER gain (\%), TC=50bp } & \multicolumn{2}{c}{Sharpe Ratio} \\[-3pt]\midrule\\[-20pt]

\multicolumn{6}{l}{\underline{Panel A: Results for $\gamma=1$}} \\[3pt]

$NR^{B}$	&	$-$1&49 	&	$-$2&29 	&	0&15 	\\
$NR^{G}$	&	5&63 	&	4&73 	&	0&51 	\\[-3pt]\midrule\\[-20pt]

\multicolumn{6}{l}{\underline{Panel B: Results for $\gamma=3$}} \\[3pt]

$NR^{B}$	&	\hspace{0.3cm}$-$3&23 	&	\hspace{1.2cm}$-$3&90 	&	\hspace{0.4cm}$-$0&07 	\\
$NR^{G}$	&	4&92 	&	3&55 	&	0&51 	\\[-3pt]\midrule\\[-20pt]

\multicolumn{6}{l}{\underline{Panel C: Results for $\gamma=5$}} \\[3pt]

$NR^{B}$	&	$-$3&21 	&	$-$3&67 	&	\hspace{0.3cm}$-$0&15 	\\
$NR^{G}$	&	2&97 	&	1&51 	&	0&53 	\\[-3pt]
\bottomrule

\end{tabular}
\end{table}

\clearpage \pagebreak

\newpage
\begin{table} [!htbp]
\captionsetup{singlelinecheck=false,justification=raggedright,labelfont=bf}
\caption{\textbf{Forecasting Macroeconomic Conditions by ChatGPT}}\vspace{-0.2cm}
{\small This table presents the results of following regression,
$$Y_{t+1}=\alpha + \beta\, NR^{K}_t + \epsilon_{t+1} \ , \qquad K=B\ \mbox{or}\ G \ ,  \vspace{-0.2cm} $$
where $Y_{t+1}$ represents macroeconomic condition variables at future time $t+1$, including the industry production growth (IPG), the VIX of CBOE, the CFNAI, the Aruoba-Diebold-Scotti Business Conditions Index (ADSI), the Kansas City Financial Stress Index (KCFSI), the total non-farm payroll growth (Payroll Growth), the smoothed recession probability, and the real GDP growth (GDPG). 
$NR^{K}$ represents news ratios. Specifically, $NR^{B}$ is the monthly proportion of bad news and $NR^{G}$ represents the proportion of good news. The good or bad news is identify by ChatGPT-3.5. It answers whether the input news means \textit{GOING UP} or \textit{GOING DOWN} for the stock market. Reported are slope estimates ($\beta$) and $R^2$s in percentage form, and also the \citet{newey1986simple} $t$-statistics. *, **, and *** indicate significance at the 10\%, 5\%, and 1\% levels, respectively. The sample period is from January 1996 to December 2022. }\label{Tab: Links to Macro}

\renewcommand{\arraystretch}{1.6}
\tabcolsep 12pt
\small
\centering
\begin{tabular}{l r@{.}lr@{.}lr@{.}l c r@{.}lr@{.}lr@{.}l}\\[-15pt]\toprule \\[-20pt]

&  \multicolumn{6}{c}{Panel A: Bad News} &&  \multicolumn{6}{c}{Panel B: Good News} \\ \cline{2-7}\cline{9-14}

&  \multicolumn{2}{c}{$\beta$}   & \multicolumn{2}{c}{NW-$t$}  & \multicolumn{2}{c}{$R^2$ (\%)} &&  \multicolumn{2}{c}{$\beta$}   & \multicolumn{2}{c}{NW-$t$}  & \multicolumn{2}{c}{$R^2$ (\%)} \\[-3pt]\midrule\\[-20pt]

IPG	&	$-$0&18$^{**}$	&	$-$1&97 	&	3&33 	&&	0&09$^{**}$	&	2&35 	&	0&89 	\\
VIX	&	0&33$^{***}$	&	3&80 	&	11&06 	&&	$-$0&14$^{**}$	&	$-$2&55 	&	2&05 	\\
CFNAI	&	$-$0&17$^{*}$	&	$-$1&69 	&	2&84 	&&	0&12$^{***}$	&	4&30 	&	1&38 	\\
ADSI	&	$-$0&13 		&	$-$1&37 	&	1&56 	&&	0&10$^{***}$	&	3&66 	&	1&00 	\\
KCFSI	&	0&42$^{***}$	&	3&30 	&	17&79 	&&	$-$0&31$^{***}$	&	$-$5&96 	&	9&77 	\\
Payroll Growth	&	$-$0&09 		&	$-$1&03 	&	0&85 	&&	0&07$^{***}$	&	4&26 	&	0&51 	\\
SRP	&	0&39$^{***}$	&	3&15 	&	15&13 	&&	$-$0&23$^{***}$	&	$-$3&64 	&	5&46 	\\
GDPG	&	$-$0&33$^{***}$	&	$-$2&84 	&	10&66 	&&	0&10$^{*}$	&	1&91 	&	0&97 	\\[-3pt] 
\bottomrule

\end{tabular}
\end{table}

\clearpage \pagebreak

\newpage
\begin{table} [!htbp]
\captionsetup{singlelinecheck=false,justification=raggedright,labelfont=bf}
\caption{\textbf{Forecasting Macroeconomic Conditions by DeepSeek}}\vspace{-0.2cm}
{\small Panel A presents the results of following regression,
$$Y_{t+1}=\alpha + \beta\, NR^{K}_t + \epsilon_{t+1} \ , \qquad K=B\ \mbox{or}\ G \ ,  \vspace{-0.2cm} $$
where $Y_{t+1}$ represents macroeconomic condition variables at future time $t+1$, including the industry production growth (IPG), the VIX of CBOE, the CFNAI, the Aruoba-Diebold-Scotti Business Conditions Index (ADSI), the Kansas City Financial Stress Index (KCFSI), the total non-farm payroll growth (Payroll Growth), the smoothed recession probability, and the real GDP growth (GDPG). 
$NR^{K}$ represents news ratios. Specifically, $NR^{B}$ is the monthly proportion of bad news and $NR^{G}$ represents the proportion of good news. The good or bad news is identify by DeepSeek-R1. It answers whether the input news means \textit{GOING UP} or \textit{GOING DOWN} for the stock market. Panel B presents the results of following regression,
$$ \Delta S_{t+1} = \alpha + \beta\, NR^{K}_{t} + \epsilon_{t+1} \ , \qquad K = B\ \mbox{or}\ G \ , $$
where $\Delta S_{t+1}$ is the monthly changes ($S_{t+1}-S_t$) in investor sentiment proposed by \citet{baker2006investor}. Reported are slope estimates ($\beta$) and $R^2$s in percentage form, and also the \citet{newey1986simple} $t$-statistics. *, **, and *** indicate significance at the 10\%, 5\%, and 1\% levels, respectively. The sample period is from January 1996 to December 2022. }\label{Tab: Links to Sentiment}

\renewcommand{\arraystretch}{1.6}
\tabcolsep 12pt
\small
\centering
\begin{tabular}{l r@{.}lr@{.}lr@{.}l c r@{.}lr@{.}lr@{.}l}\\[-15pt]\toprule \\[-20pt]

&  \multicolumn{6}{c}{\emph{Bad} News Ratio ($NR^B$)} &&  \multicolumn{6}{c}{\emph{Good} News Ratio ($NR^G$)} \\ \cline{2-7}\cline{9-14}

&  \multicolumn{2}{c}{$\beta$}   & \multicolumn{2}{c}{NW-$t$}  & \multicolumn{2}{c}{$R^2$ (\%)} &&  \multicolumn{2}{c}{$\beta$}   & \multicolumn{2}{c}{NW-$t$}  & \multicolumn{2}{c}{$R^2$ (\%)} \\[-3pt]\midrule\\[-20pt]

\multicolumn{14}{l}{\underline{Panel A: Forecasting Macroeconomy}}\\[3pt]

IPG	&	$-$0&22$^{**}$	&	$-$1&96 	&	4&67 		&&	0&04 		&	0&89 	&	0&17 	\\
VIX	&	0&39$^{***}$	&	4&76 	&	14&89 		&&	$-$0&04 		&	$-$0&62 	&	0&14 	\\
CFNAI	&	$-$0&21$^{*}$	&	$-$1&75 	&	4&46 		&&	0&05 		&	1&30 	&	0&26 	\\
ADSI	&	$-$0&16 		&	$-$1&58 	&	2&69 		&&	0&05 		&	1&38 	&	0&22 	\\
KCFSI	&	0&46$^{***}$	&	4&25 	&	21&53 		&&	$-$0&15$^{***}$	&	$-$2&70 	&	2&28 	\\
Payroll Growth	&	$-$0&15 		&	$-$1&16 	&	2&13 		&&	0&02 		&	0&44 	&	0&02 	\\
SRP	&	0&45$^{***}$	&	3&84 	&	20&01 		&&	$-$0&10$^{**}$	&	$-$2&10 	&	1&02 	\\
GDPG	&	$-$0&37$^{***}$	&	$-$2&84 	&	13&39 		&&	0&01 		&	0&20 	&	0&02 	\\[-3pt]\midrule\\[-20pt] 

\multicolumn{14}{l}{\underline{Panel B: Forecasting Sentiment}}\\[3pt]

ChatGPT	&	$-$0&18$^{***}$ 	&	$-$2&72 	&	3&11 		&&	0&10 	&	1&14 	&	0&91 	\\
DeepSeek	&	$-$0&20$^{***}$ 	&	$-$3&66 	&	4&13 		&&	0&12$^{*}$ 	&	1&67 	&	1&37 	\\[-3pt]
\bottomrule

\end{tabular}
\end{table}

\clearpage \pagebreak

\newpage
\begin{landscape}
\begin{table} [!htbp]
\captionsetup{singlelinecheck=false,justification=raggedright,labelfont=bf}
\caption{\textbf{Relations to the SPF Expectation}}\vspace{-0.2cm}
{\small This table presents the results of following regression,
$$ E_{t} = \alpha + \beta\, NR^{K}_{t-1} + \psi\, E_{t-1} + \epsilon_{t} \ , \qquad K = B\ \mbox{or}\ G \ , $$
where $E_{t}$ is the equal-weighted quarterly forecasts on six economic variables: the real GDP growth, industrial production growth, unemployment rate, non-farm payroll growth, T-bill yield, and the inflation rate obtained from the Survey of Professional Forecasters (SPF). In each quarter, SPF provides the forecasts for current quarter (nowcasting) and subsequent four quarters. To make the values of the forecasts comparable, we standardize them before equal weighting. \( NR^{K} \) represents the quarterly news ratios, either good news ratio \( NR^{G} \) or bad news ratio \( NR^{B} \), defined as the proportion of good or bad news within each quarter. The good or bad news is identified by ChatGPT-3.5. Reported are slope estimates and $R^2$s in percentage form, and also the \citet{newey1986simple} $t$-statistics. *, **, and *** indicate significance at the 10\%, 5\%, and 1\% levels, respectively. The sample period spans from 1996 to 2022. }\label{Tab: Macro Expectations}

\renewcommand{\arraystretch}{1.6}
\tabcolsep 12pt
\small
\centering
\begin{tabular}{l r@{.}lr@{.}lr@{.}lr@{.}lr@{.}l c r@{.}lr@{.}lr@{.}lr@{.}lr@{.}l}\\[-15pt]\toprule \\[-20pt]

&  \multicolumn{10}{c}{Panel A: Bad News} &&  \multicolumn{10}{c}{Panel B: Good News} \\ \cline{2-11}\cline{13-22}

&  \multicolumn{2}{c}{$\beta$}   & \multicolumn{2}{c}{NW-$t$}  &  \multicolumn{2}{c}{$\psi$}   & \multicolumn{2}{c}{NW-$t$} & \multicolumn{2}{c}{$R^2$ (\%)} &&  \multicolumn{2}{c}{$\beta$}   & \multicolumn{2}{c}{NW-$t$}  &  \multicolumn{2}{c}{$\psi$}   & \multicolumn{2}{c}{NW-$t$}  & \multicolumn{2}{c}{$R^2$ (\%)} \\[-3pt]\midrule\\[-20pt]

Nowcasting	&	$-$0&20$^{**}$	&	$-$2&46 	&	$-$0&14 		&	$-$0&60 	&	9&50 	&&	0&04 		&	0&92 	&	0&06 		&	0&25 	&	0&92 	\\
1 quarter	&	$-$0&09$^{***}$	&	$-$3&10 	&	0&54$^{***}$	&	8&06 	&	38&04 	&&	0&06$^{*}$	&	1&70 	&	0&59$^{***}$	&	4&54 	&	36&69 	\\
2 quarter	&	$-$0&09$^{***}$	&	$-$4&10 	&	0&56$^{***}$	&	6&84 	&	42&21 	&&	0&03 		&	0&95 	&	0&63$^{***}$	&	4&44 	&	39&58 	\\
3 quarter	&	$-$0&09$^{***}$	&	$-$3&42 	&	0&61$^{***}$	&	4&89 	&	46&74 	&&	$-$0&01 		&	$-$0&36 	&	0&67$^{***}$	&	4&18 	&	43&85 	\\
4 quarter	&	$-$0&05 		&	$-$1&62 	&	0&70$^{***}$	&	4&82 	&	53&43 	&&	$-$0&04 		&	$-$1&06 	&	0&71$^{***}$	&	4&42 	&	52&92 	\\[-3pt] 
\bottomrule

\end{tabular}
\end{table}
\end{landscape}

\clearpage \pagebreak

\normalsize\newpage
\begin{landscape}
\begin{table} [!htbp]
\captionsetup{singlelinecheck=false,justification=raggedright,labelfont=bf}
\caption{\textbf{Interaction between Good News Ratio and Economic Activity }}\vspace{-0.2cm}
{\small This table reports estimation results of the following regression,
$$ R_{t+h} = \alpha + \beta_1\,I_{High}\, NR^G_t  + \beta_2\,I_{Low}\,NR^G_t + \beta_3\, I_{High} + \epsilon_{t+h} \ ,  \vspace{-0.1cm} $$
where $R_{t+h}$ is the average excess return of market portfolio from $t+1$ to $t+h$, $h=$ 1, 3, 6, 9, and 12 months. $NR^G$ is the monthly proportion of good news identified by ChatGPT-3.5. It answers whether the input news means \textit{GOING UP} for the stock market. $I_{High}$ is an indicator variable which equals one if current Chicago Fed National Activity Index (CFNAI) exceeds the past five-year sample mean and zero otherwise. The counterpart variable $I_{Low}$ equals to $1-I_{High}$. All forecasting variables are standardized to have a zero mean and unit variance. Reported are the regression slopes and $R^2$s in percentage form. Brackets below the slope estimates report the \citet{hodrick1992dividend} $t$-statistics. *, **, and *** indicate significance at the 10\%, 5\%, and 1\% levels, respectively. The sample period is from January 1996 to December 2022. } \label{Tab: Econ Activity}

\renewcommand{\arraystretch}{1.6}
\tabcolsep 33pt
\small
\centering

\begin{tabular}{l r@{.}lr@{.}lr@{.}lr@{.}lr@{.}l} \\[-15pt]\toprule\\[-25pt]

& \multicolumn{2}{c}{$h=1$}    & \multicolumn{2}{c}{\hspace{-0.2cm}$h=3$}    &  \multicolumn{2}{c}{$h=6$}    & \multicolumn{2}{c}{$h=9$}   & \multicolumn{2}{c}{$h=12$}  \\[-3pt] \midrule \\[-17pt]

$I_{High}\times NR^G$	&	0&12 	&	0&17 	&	$-$0&03 	&	$-$0&08 	&	$-$0&06 	\\
	&	[0&44] 	&	[0&68] 	&	[$-$0&12] 	&	[$-$0&38] 	&	[$-$0&38] 	\\
$I_{Low}\times NR^G$	&	0&71$^{*}$ 	&	0&83$^{***}$ 	&	0&97$^{***}$ 	&	0&93$^{***}$ 	&	0&84$^{***}$ 	\\
	&	[1&86] 	&	[2&91] 	&	[3&35] 	&	[3&08] 	&	[3&08] 	\\
$I_{High}$	&	0&90 	&	0&55 	&	0&53$^{*}$ 	&	0&42 	&	0&44$^{*}$ 	\\
	&	[1&58] 	&	[1&49] 	&	[1&73] 	&	[1&40] 	&	[1&74] 	\\
$R^2$ (\%)	&	1&79 	&	6&36 	&	14&42 	&	17&64 	&	19&64 	\\[-3pt]
\bottomrule

\end{tabular}
\end{table}
\end{landscape}

\clearpage \pagebreak

\normalsize\newpage
\begin{landscape}
\begin{table} [!htbp]
\captionsetup{singlelinecheck=false,justification=raggedright,labelfont=bf}
\caption{\textbf{Interaction between Good News Ratio and EPU }} \vspace{-0.2cm}
{\small This table reports estimation results of the following regression,
$$ R_{t+h} = \alpha + \beta_1\,U_{High}\, NR^G_t  + \beta_2\,U_{Low}\,NR^G_t + \beta_3\, U_{High} + \epsilon_{t+h} \ ,  \vspace{-0.1cm} $$
where $R_{t+h}$ is the average excess return of market portfolio from $t+1$ to $t+h$, $h=$ 1, 3, 6, 9, and 12 months. $NR^G$ is the monthly proportion of good news identified by ChatGPT-3.5. It answers whether the input news means \textit{GOING UP} for the stock market. $U_{High}$ is an indicator variable which equals one if current Economic Policy Uncertainty (EPU) exceeds the past five-year sample mean and zero otherwise. The counterpart variable $U_{Low}$ equals to $1-U_{High}$. All forecasting variables are standardized to have a zero mean and unit variance. Reported are the regression slopes and $R^2$s in percentage form. Brackets below the slope estimates report the \citet{hodrick1992dividend} $t$-statistics. *, **, and *** indicate significance at the 10\%, 5\%, and 1\% levels, respectively. The sample period is from January 1996 to December 2022. } \label{Tab: EPU}

\renewcommand{\arraystretch}{1.6}
\tabcolsep 33pt
\small
\centering

\begin{tabular}{l r@{.}lr@{.}lr@{.}lr@{.}lr@{.}l} \\[-15pt]\toprule\\[-25pt]

& \multicolumn{2}{c}{$h=1$}    & \multicolumn{2}{c}{\hspace{-0.2cm}$h=3$}    &  \multicolumn{2}{c}{\hspace{-0.2cm}$h=6$}    & \multicolumn{2}{c}{$h=9$}   & \multicolumn{2}{c}{$h=12$}  \\[-3pt] \midrule \\[-17pt]

$U_{High}\times NR^G$	&	0&78$^{**}$ 	&	0&81$^{***}$ 	&	0&84$^{***}$ 	&	0&89$^{***}$ 	&	0&85$^{***}$ 	\\
	&	[2&20] 	&	[3&11] 	&	[2&90] 	&	[3&02] 	&	[2&90] 	\\
$U_{Low}\times NR^G$	&	0&26 	&	0&29 	&	0&15 	&	$-$0&05 	&	$-$0&05 	\\
	&	[0&99]	&	[1&07] 	&	[0&59] 	&	[$-$0&23] 	&	[$-$0&25] 	\\
$U_{High}$	&	0&13 	&	0&26 	&	0&02 	&	$-$0&07 	&	0&09 	\\
	&	[0&28] 	&	[0&72] 	&	[0&05] 	&	[$-$0&23] 	&	[0&39] 	\\
$R^2$ (\%)	&	0&79 	&	4&95 	&	9&35 	&	15&15 	&	17&77 	\\[-3pt]
\bottomrule

\end{tabular}
\end{table}
\end{landscape}

\clearpage \pagebreak

\normalsize\newpage
\begin{landscape}
\begin{table} [!htbp]
\captionsetup{singlelinecheck=false,justification=raggedright,labelfont=bf}
\caption{\textbf{Interaction between Good News Ratio and News Similarity}}\vspace{-0.2cm}
{\small This table reports estimation results of the following regression,
$$ R_{t+h} = \alpha + \beta_1\,S_{High}\, NR^G_t  + \beta_2\,S_{Low}\,NR^G_t + \beta_3\, S_{High} + \epsilon_{t+h} \ ,  \vspace{-0.1cm} $$
where $R_{t+h}$ is the average excess return of market portfolio from $t+1$ to $t+h$, $h=$ 1, 3, 6, 9, and 12 months. $NR^G$ is the monthly proportion of good news identified by ChatGPT-3.5. It answers whether the input news means \textit{GOING UP} for the stock market. \( S_{High} \) is an indicator variable which equals one if current similarity of economic news exceeds the past five-year sample mean and zero otherwise. The counterpart variable $S_{Low}$ equals to $1 - S_{High}$. The economic news refers to news that contain economic-relevant keywords listed in Appendix B. All forecasting variables are standardized to have a zero mean and unit variance. Reported are the regression slopes and \( R^2 \)s in percentage form. Brackets below the slope estimates report the \citet{hodrick1992dividend} $t$-statistics. *, **, and *** indicate significance at the 10\%, 5\%, and 1\% levels, respectively. The sample period is from January 1996 to December 2022.}
\label{Tab: Similarity}

\renewcommand{\arraystretch}{1.6}
\tabcolsep 33pt
\small
\centering

\begin{tabular}{l r@{.}lr@{.}lr@{.}lr@{.}lr@{.}l} \\[-15pt]\toprule\\[-25pt]

& \multicolumn{2}{c}{$h=1$}    & \multicolumn{2}{c}{$h=3$}    &  \multicolumn{2}{c}{\hspace{-0.2cm}$h=6$}    & \multicolumn{2}{c}{$h=9$}   & \multicolumn{2}{c}{$h=12$}  \\[-3pt] \midrule \\[-17pt]

\( S_{High} \times NR^G \) & 0&24 & 0&44 & 0&34 & 0&26 & 0&22 \\
& [0&75] & [1&37] & [0&88] & [0&72] & [0&71] \\
\( S_{Low} \times NR^G \) & 0&84\( ^{**} \) & 0&71\( ^{**} \) & 0&71\( ^{***} \) & 0&67\( ^{***} \) & 0&68\( ^{***} \) \\
& [2&35] & [2&19] & [2&97] & [2&83] & [3&04] \\
\( S_{High} \) & 0&65 & $-$0&12 & 0&02 & $-$0&12 & $-$0&20 \\
& [1&19] & [$-$0&27] & [0&06] & [$-$0&41] & [$-$1&35] \\
\( R^2 (\%) \) & 1&38 & 4&02 & 6&96 & 8&23 & 10&61 \\[-3pt]
\bottomrule

\end{tabular}
\end{table}
\end{landscape}

\clearpage \pagebreak

\setstretch{1.6}

\begin{center}\label{internet_appendix}
 \LARGE\textbf{\\ Internet Appendix for \\[1.5ex] ChatGPT and DeepSeek: Can They Predict the Stock Market and Macroeconomy?}
 \end{center}

\date{\vspace{0.5cm}First Draft: July 2023\\[1.2ex] Current Draft: December 2023}


\thispagestyle{empty}

\vspace{2cm}\setlength{\parindent}{0.5em}
This Internet Appendix reports the results for supplementary and robustness tests:\\
\begin{center}
\begin{tabular}{lp{14cm}} \\[-30pt]
\textbf{Figure IA 1:} & Weekly Out-of-sample Results \\
\textbf{Table IA 1:} & Robust Check for Table 2 \\
\textbf{Table IA 2:} & Comparisons with Alternative LLMs \\
\textbf{Table IA 3:} & Comparisons with Economic Variables\\
\textbf{Table IA 4:} & Control for Lagged Return\\
\textbf{Table IA 5:} & Additional Results in Table 4\\
\textbf{Table IA 6:} & Supplementary Results in Table 8\\

\end{tabular}
\end{center}

\thispagestyle{empty}
\newpage
\baselineskip=20pt
\setcounter{page}{1} \setcounter{equation}{0}
\setstretch{1.6}

\normalsize\newpage

\begin{figure}[H] 
\textbf{Figure IA 1. Weekly Out-of-sample Forecasting Errors}\\[3pt]
{\small This figure plots the weekly difference between the cumulative squared forecast error (CSFE) generated by the historical average benchmark forecast and the CSFE derived from forecasts based on good news ratio ($NR^{G}$) or bad news ratio ($NR^{B}$), which are the past four-week proportion of good or bad news identified by ChatGPT-3.5. It answers whether the input news means \textit{GOING UP} or \textit{GOING Down} for stock market. The full data sample period spans from October 2021 to December 2023, in which the first 35 weeks are used as training sample and the remaining weeks are out-of-sample evaluation sample. }

\begin{center}
\includegraphics[viewport=1cm 1cm 20cm 23cm]{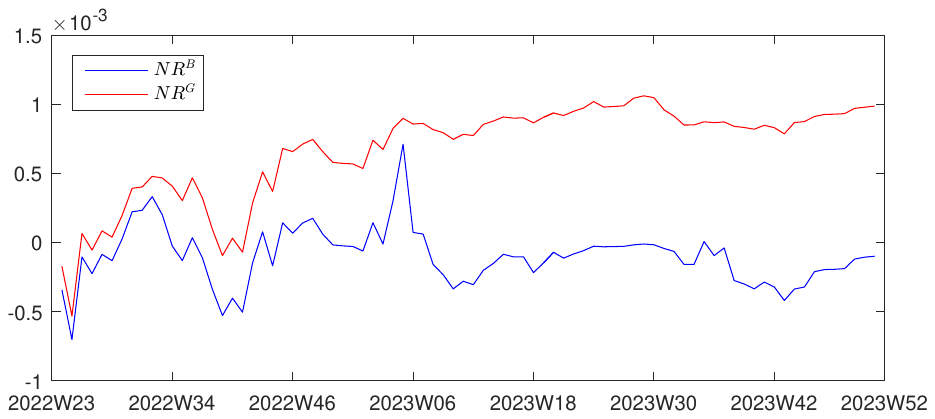}
\end{center}

\end{figure}








\clearpage \pagebreak


\normalsize\newpage
\begin{table} [!htbp]
\begin{threeparttable}  
\textbf{Table IA 1. Results for Alternative Statistical Standard Errors }\\[3pt]
{\small This table reports estimation results of the following regression,\vspace{-0.1cm}
$$ R_{t+h} = \alpha + \beta\, NR_{t}^{K} + \epsilon_{t+h} \ , \qquad K=B\ \mbox{or}\ G \ ,  \vspace{-0.2cm} $$
where $R_{t}$ denotes the current market excess return on the S\&P 500 index at time $t$ for "$h=0$". This setting aligns with a contemporaneous regression framework. For scenarios where $h>0$, $R_{t+h}$ represents the average excess returns of the market portfolio from $t+1$ to $t+h$ (with $h$ being 1, 3, 6, 9, and 12 months), transitioning the equation into a predictive regression. $NR^{K}$ represents news ratios. Specifically, $NR^{B}$ is the monthly proportion of bad news and $NR^{G}$ represents the proportion of good news. The good or bad news is identified by ChatGPT-3.5. It answers whether the input news means \textit{GOING UP} or \textit{GOING DOWN} for the stock market. Reported are the regression slopes and $R^2$s in percentage form. Also reported are the \citet{newey1986simple} $t$-statistics (NW-$t$). All the forecast variables are standardized to have a zero mean and unit variance. *, **, and *** indicate significance at the 10\%, 5\%, and 1\% levels, respectively. The sample period is from January 1996 to December 2022.  } \end{threeparttable}

\renewcommand{\arraystretch}{1.6}
\tabcolsep 40pt
\small
\centering

\begin{tabular}{l r@{.}lr@{.}lr@{.}l}\\[-15pt] \toprule\\[-20pt]

& \multicolumn{2}{c}{$\beta$ (\%)}   & \multicolumn{2}{c}{\hspace{0.3cm}NW-$t$}        & \multicolumn{2}{c}{$R^2$ (\%)}\\[-3pt] \midrule \\[-20pt]

\multicolumn{7}{l}{\underline{Panel A: Results for $NR^{B}$}}\\

$h=$ 0	&	$-$1&03$^{***}$	&	\hspace{0.1cm}$-$2&77 	&	\hspace{0.1cm}5&17 	\\	
$h=$ 1	&	0&05 		&	0&14 	&	0&01 	\\
$h=$ 3	&	0&06 		&	0&21 	&	0&05 	\\
$h=$ 6	&	0&14 		&	0&54 	&	0&48 	\\
$h=$ 9	&	0&21 		&	0&85 	&	1&57 	\\
$h=$ 12	&	0&25 		&	1&03 	&	2&65 	\\[-3pt]\midrule\\[-15pt]
								
\multicolumn{7}{l}{\underline{Panel B: Results for $NR^{G}$}}\\								
								
$h=$ 0	&	1&02$^{***}$	&	4&78 	&	5&07 	\\	
$h=$ 1	&	0&53$^{**}$	&	2&21 	&	1&37 	\\	
$h=$ 3	&	0&56$^{**}$	&	2&66 	&	4&60 	\\	
$h=$ 6	&	0&51$^{*}$	&	2&21 	&	6&91 	\\	
$h=$ 9	&	0&44$^{*}$ 		&	1&83 	&	7&46 	\\
$h=$ 12	&	0&42$^{*}$ 		&	1&74 	&	8&52 	\\[-3pt]
\bottomrule

\end{tabular}
\end{table}

\clearpage \pagebreak

\begin{table} [!htbp]
\begin{threeparttable}
\textbf{Table IA 2. Forecasting Results for RoBERTa}\\[3pt]
{\small This table reports estimation results of the following regression,
$$ R_{t+h} = \alpha + \beta\, NR_{t}^{K} + \epsilon_{t+h} \ , \qquad K=B\ \mbox{or}\ G \ ,  \vspace{-0.2cm} $$
where $R_{t}$ denotes the current market excess return on the S\&P 500 index at time $t$ for "$h=0$". This setting aligns with a contemporaneous regression framework. For scenarios where $h>0$, $R_{t+h}$ represents the average excess returns of the market portfolio from $t+1$ to $t+h$ (with $h$ being 1, 3, 6, 9, and 12 months), transitioning the equation into a predictive regression. $NR^{K}$ represents news ratios. Specifically, $NR^{B}$ is the monthly proportion of bad news and $NR^{G}$ represents the proportion of good news. The good or bad news is identified by RoBERTa. It answers whether the input news means \textit{GOING UP} or \textit{GOING DOWN} for the stock market.  Reported are the regression slopes and $R^2$s in percentage form. Also reported are the \citet{hodrick1992dividend} $t$-statistics. All the forecast variables are standardized to have a zero mean and unit variance. *, **, and *** indicate significance at the 10\%, 5\%, and 1\% levels, respectively. The sample period is from January 1996 to December 2022.}
\end{threeparttable}

\renewcommand{\arraystretch}{1.6}
\tabcolsep 14pt
\small
\centering

\begin{tabular}{l r@{.}lr@{.}lr@{.}l c r@{.}lr@{.}lr@{.}l}\\[-15pt] \toprule\\[-20pt]

& \multicolumn{6}{c}{Panel A: Results for $NR^{B}$}  && \multicolumn{6}{c}{Panel B: Results for $NR^{G}$} \\ \cline{2-7}\cline{9-14} \\[-20pt]

& \multicolumn{2}{c}{$\beta$ (\%)}   & \multicolumn{2}{c}{Hodrick-$t$}        & \multicolumn{2}{c}{$R^2$ (\%)} && \multicolumn{2}{c}{$\beta$ (\%)}   & \multicolumn{2}{c}{Hodrick-$t$}        & \multicolumn{2}{c}{$R^2$ (\%)} \\[-3pt] \midrule \\[-20pt]

\(h=0\)	&	$-$0&61\(^{*}\)	&	\hspace{0.1cm}$-$1&94 	&	1&84 	&&	$-$0&16 		&	\hspace{0.1cm}$-$0&63 	&	0&12 	\\
\(h=1\)	&	0&36 		&	1&32 	&	0&64 	&&	0&17 		&	0&63 	&	0&13 	\\
\(h=3\)	&	0&24 		&	0&95 	&	0&86 	&&	0&32\(^{*}\)	&	1&81 	&	1&44 	\\
\(h=6\)	&	0&22 		&	0&89 	&	1&24 	&&	0&25 		&	1&60 	&	1&62 	\\
\(h=9\)	&	0&16 		&	0&62 	&	0&89 	&&	0&16 		&	1&02 	&	0&90 	\\
\(h=12\)	&	0&20 		&	0&78 	&	1&63 	&&	0&17 		&	1&26 	&	1&29 	\\[-3pt]
\bottomrule

\end{tabular}
\end{table}

\clearpage \pagebreak

\begin{landscape}
\begin{table} [!htbp]
\begin{threeparttable}
\textbf{Table IA 3. Comparisons with Economic Variables}\\[3pt]
{\small This table reports estimation results of the following regression,
$$ R_{t+h} = \alpha + \beta_1\, NR_{t}^{G} + \beta_2\, PC1_{t} + \beta_3\, PC2_{t} + \beta_4\,PC3_{t} + \beta_5\, PC4_{t} + + \beta_6\, PC5_{t}  + \epsilon_{t+h} \ ,   \vspace{-0.2cm} $$
where $R_{t+h}$ represents the average excess returns of the market portfolio from $t+1$ to $t+h$ (with $h$ being 1, 3, 6, 9, and 12 months); $NR^{G}$ represents the proportion of good news identified by ChatGPT-3.5. It answers whether the input news means \textit{GOING UP} for the stock market; PC1--PC5 are the first five principal components of the 14 economic variables proposed by \citet{welch2008comprehensive}. Reported are the regression slopes and \( R^2 \)s in percentage form. Brackets below the slope estimates report the \citet{hodrick1992dividend} $t$-statistics. All the forecast variables are standardized to have a zero mean and unit variance. *, **, and *** indicate significance at the 10\%, 5\%, and 1\% levels, respectively. The sample period is from January 1996 to December 2022.}
\end{threeparttable}

\renewcommand{\arraystretch}{1.6}
\tabcolsep 25pt
\small
\centering

\begin{tabular}{l r@{.}lr@{.}lr@{.}lr@{.}lr@{.}lr@{.}lr@{.}l}\\[-15pt] \toprule\\[-20pt]

& \multicolumn{2}{c}{$\beta_1$}   & \multicolumn{2}{c}{$\beta_2$}        & \multicolumn{2}{c}{$\beta_3$} & \multicolumn{2}{c}{$\beta_4$}  & \multicolumn{2}{c}{$\beta_5$} & \multicolumn{2}{c}{$\beta_6$} & \multicolumn{2}{c}{$R^2$ (\%)} \\[-3pt] \midrule \\[-20pt]

$h=$ 1	&	0&58$^{**}$	&	$-$0&16 		&	0&22 	&	0&03 	&	$-$0&24 	&	0&27 		&	2&36 	\\
	&	[2&04] 		&	[$-$0&39] 		&	[0&60] 	&	[0&06] 	&	[$-$0&95] 	&	[0&92] 		&	\multicolumn{2}{c}{}	\\
$h=$ 3	&	0&78$^{***}$	&	$-$0&17 		&	0&03 	&	0&10 	&	$-$0&27 	&	0&52$^{**}$	&	9&56 	\\
	&	[2&99] 		&	[$-$0&43] 		&	[0&11] 	&	[0&26] 	&	[$-$1&30] 	&	[2&06] 		&	\multicolumn{2}{c}{}	\\
$h=$ 6	&	0&76$^{**}$	&	$-$0&27 		&	$-$0&04 	&	0&18 	&	$-$0&33 	&	0&54$^{**}$	&	18&76 	\\
	&	[2&51] 		&	[$-$0&96] 		&	[$-$0&26] 	&	[1&04] 	&	[$-$1&46] 	&	[2&50] 		&	\multicolumn{2}{c}{}	\\
$h=$ 9	&	0&65$^{**}$	&	$-$0&34 		&	$-$0&01 	&	0&18 	&	$-$0&34 	&	0&51$^{**}$	&	24&91 	\\
	&	[2&11] 		&	[$-$1&60] 		&	[$-$0&05] 	&	[1&38] 	&	[$-$1&42] 	&	[2&44] 		&	\multicolumn{2}{c}{}	\\
$h=$ 12	&	0&55$^{**}$	&	$-$0&39$^{**}$	&	0&04 	&	0&13 	&	$-$0&34 	&	0&45$^{**}$	&	29&65 	\\
	&	[2&15] 		&	[$-$2&22] 		&	[0&22] 	&	[1&20] 	&	[$-$1&39] 	&	[2&46] 		&	\multicolumn{2}{c}{}	\\[-3pt]
\bottomrule

\end{tabular}
\end{table}
\end{landscape}

\clearpage \pagebreak


\normalsize\newpage
\begin{table} [!htbp]
\begin{threeparttable}  
\textbf{Table IA 4. Control for Lagged Return }\\[3pt]
{\small This table reports estimation results of the following regression,\vspace{-0.1cm}
$$ R_{t+h} = \alpha + \beta\, NR_{t}^{G} + \psi\, R_{t} + \epsilon_{t+h} \ ,  \vspace{-0.2cm} $$
where $R_{t+h}$ is the average excess returns of market portfolio from $t+1$ to $t+h$, $h=$ 1, 3, 6, 9, and 12 months.  $NR^{G}$ is the monthly proportion of good news identified by ChatGPT-3.5. It answers whether the input news means \textit{GOING UP} for the stock market. $R_t$ is the current market excess return at time $t$, which is added as a control variable. Reported are the regression slopes and $R^2$s in percentage form. Also reported are the \citet{hodrick1992dividend} $t$-statistics. All the forecast variables are standardized to have a zero mean and unit variance. *, **, and *** indicate significance at the 10\%, 5\%, and 1\% levels, respectively. The sample period is from January 1996 to December 2022.  } \end{threeparttable}

\renewcommand{\arraystretch}{1.6}
\tabcolsep 22pt
\small
\centering

\begin{tabular}{l r@{.}lr@{.}lr@{.}lr@{.}lr@{.}l}\\[-15pt] \toprule\\[-20pt]

& \multicolumn{2}{c}{$\beta$ (\%)}   & \multicolumn{2}{c}{Hodrick-$t$}        &  \multicolumn{2}{c}{\hspace{0.2cm}$\psi$ (\%)}  & \multicolumn{2}{c}{Hodrick-$t$}  & \multicolumn{2}{c}{$R^2$ (\%)}\\[-3pt] \midrule \\[-20pt]
								
$h=1$	&	0&54$^{**}$ 	&	\hspace{0.3cm}2&27	&	$-$0&03 	&	\hspace{0.2cm}$-$0&12 	&	1&37 	\\
$h=3$	&	0&57$^{**}$ 	&	2&50	&	$-$0&06 	&	$-$0&33 	&	4&65 	\\
$h=6$	&	0&52$^{*}$ 	&	1&90	&	$-$0&06 	&	$-$0&68 	&	7&01 	\\
$h=9$	&	0&44 	&	1&50 		&	$-$0&02 	&	$-$0&23 	&	7&47 	\\
$h=12$	&	0&42 	&	1&49 		&	$-$0&02 	&	$-$0&33 	&	8&54 	\\[-3pt]
\bottomrule

\end{tabular}
\end{table}

\clearpage \pagebreak

\begin{table} [!htbp]
\begin{threeparttable}
\textbf{Table IA 5. Additional Results of Alternative Prompts}\\[3pt]
{\small This table reports estimation results of the following regression,
$$ R_{t+h} = \alpha + \beta\,NR_{t}^K + \epsilon_{t+h}\ , \qquad K=B\ \mbox{or}\ G \ ,  \vspace{-0.2cm} $$
where $R_{t}$ denotes the current market excess return on the S\&P 500 index at time $t$ for "$h=0$". This setting aligns with a contemporaneous regression framework. For scenarios where $h>0$, $R_{t+h}$ represents the average excess returns of the market portfolio from $t+1$ to $t+h$ (with $h$ being 1, 3, 6, 9, and 12 months), transitioning the equation into a predictive regression. $NR^{K}$ represents news ratios. Specifically, $NR^{B}$ is the monthly proportion of bad news and $NR^{G}$ represents the proportion of good news. The good or bad news is identified by ChatGPT-3.5. It answers whether the input news is \textit{GOOD} or \textit{BAD} for the stock market. Reported are the regression slopes and \( R^2 \)s in percentage form. Also reported are the \citet{hodrick1992dividend} $t$-statistics. All the forecast variables are standardized to have a zero mean and unit variance. *, **, and *** indicate significance at the 10\%, 5\%, and 1\% levels, respectively. The sample period is from January 1996 to December 2022.}
\end{threeparttable}

\renewcommand{\arraystretch}{1.6}
\tabcolsep 14pt
\small
\centering

\begin{tabular}{l r@{.}lr@{.}lr@{.}l c r@{.}lr@{.}lr@{.}l}\\[-15pt] \toprule\\[-20pt]

& \multicolumn{6}{c}{Panel A: Results for \textit{Bad} News}  && \multicolumn{6}{c}{Panel B: Results for \textit{Good} News} \\ \cline{2-7}\cline{9-14} \\[-20pt]

& \multicolumn{2}{c}{$\beta$ (\%)}   & \multicolumn{2}{c}{Hodrick-$t$}        & \multicolumn{2}{c}{$R^2$ (\%)} && \multicolumn{2}{c}{$\beta$ (\%)}   & \multicolumn{2}{c}{Hodrick-$t$}        & \multicolumn{2}{c}{$R^2$ (\%)} \\[-3pt] \midrule \\[-20pt]

\(h=0\) & $-$0&58\(^{*}\) & \hspace{0.1cm}$-$1&82 & 1&63 && 0&68\(^{***}\) & \hspace{0.4cm}2&98 & 2&28 \\
\(h=1\) & $-$0&07 & $-$0&26 & 0&03 && 0&44\(^{*}\) & 1&94 & 0&96 \\
\(h=3\) & 0&06 & 0&21 & 0&05 && 0&46\(^{**}\) & 2&12 & 3&09 \\
\(h=6\) & 0&16 & 0&63 & 0&65 && 0&51\(^{**}\) & 2&01 & 7&21 \\
\(h=9\) & 0&25 & 0&98 & 2&26 && 0&50\(^{*}\) & 1&71 & 9&82 \\
\(h=12\) & 0&33 & 1&25 & 5&08 && 0&49 & 1&61 & 12&12 \\[-3pt]
\bottomrule

\end{tabular}
\end{table}

\clearpage\pagebreak

\newpage
\begin{table} [!htbp]
\begin{threeparttable} 
\textbf{Table IA 6. Forecasting Macroeconomic Conditions}\\[3pt]
{\small This table presents the results of following regression,
$$Y_{t+1}=\alpha + \beta\, NR^{K}_t + \epsilon_{t+1} \ , \qquad K=B\ \mbox{or}\ G \ ,  \vspace{-0.2cm} $$
where $Y_{t+1}$ represents macroeconomic condition variables at future time $t+1$, including the industry production growth (IPG), the VIX of CBOE, the CFNAI, the Aruoba-Diebold-Scotti Business Conditions Index (ADSI), the Kansas City Financial Stress Index (KCFSI), the total non-farm payroll growth (Payroll Growth), the smoothed recession probability, and the real GDP growth (GDPG). $NR^{K}$ represents news ratios. Specifically, $NR^{B}$ is monthly proportion of bad news and $NR^{G}$ is monthly proportion of good news. We identify the good or bad news by using the method of word lists proposed by \citet*{loughran2011liability} or using BERT. Reported are slope estimates ($\beta$) and $R^2$s in percentage form, and also the \citet{newey1986simple} $t$-statistics. *, **, and *** indicate significance at the 10\%, 5\%, and 1\% levels, respectively. The sample period is from January 1996 to December 2022. }
\end{threeparttable} 

\renewcommand{\arraystretch}{1.6}
\tabcolsep 12pt
\small
\centering
\begin{tabular}{l r@{.}lr@{.}lr@{.}l c r@{.}lr@{.}lr@{.}l}\\[-15pt]\toprule \\[-20pt]

&  \multicolumn{6}{c}{Panel A: Bad News} &&  \multicolumn{6}{c}{Panel B: Good News} \\ \cline{2-7}\cline{9-14}

&  \multicolumn{2}{c}{$\beta$}   & \multicolumn{2}{c}{\hspace{0.2cm}NW-$t$}  & \multicolumn{2}{c}{$R^2$ (\%)} &&  \multicolumn{2}{c}{$\beta$}   & \multicolumn{2}{c}{\hspace{0.2cm}NW-$t$}  & \multicolumn{2}{c}{$R^2$ (\%)} \\[-3pt]\midrule\\[-20pt]

\multicolumn{13}{l}{\underline{\textbf{Word Lists}}}\\[3pt]

IPG	&	$-$0&18 	$^{**}$	&	$-$2&03 	&	\hspace{0.2cm}3&20 	&&	0&10 		&	1&50 	&	\hspace{0.2cm}1&07 	\\
VIX	&	0&14 		&	1&58 	&	2&06 	&&	$-$0&10$^{*}$	&	$-$1&66 	&	0&93 	\\
CFNAI	&	$-$0&17$^{*}$	&	$-$1&83 	&	2&93 	&&	0&12$^{*}$	&	1&65 	&	1&39 	\\
ADSI	&	$-$0&17$^{*}$	&	$-$1&73 	&	2&94 	&&	0&08$^{*}$	&	1&88 	&	0&57 	\\
KCFSI	&	0&20$^{*}$	&	1&89 	&	4&07 	&&	$-$0&17$^{***}$	&	$-$2&62 	&	2&80 	\\
Payroll Growth	&	$-$0&12 		&	$-$1&19 	&	1&36 	&&	0&11 		&	1&31 	&	1&11 	\\
SRP	&	0&27$^{***}$	&	2&70 	&	7&10 	&&	$-$0&13$^{**}$	&	$-$2&00 	&	1&82 	\\
GDPG	&	$-$0&21$^{**}$	&	$-$2&27 	&	4&27 	&&	0&17$^{***}$	&	2&88 	&	2&83 	\\[-3pt]\midrule\\[-20pt]

\multicolumn{13}{l}{\underline{\textbf{Bert}}}\\[3pt]

IPG	&	$-$0&12 		&	$-$1&39 	&	1&45 	&&	$-$0&05 		&	$-$0&86 	&	0&22 	\\
VIX	&	0&12 		&	1&52 	&	1&47 	&&	0&17$^{**}$	&	2&40 	&	3&02 	\\
CFNAI	&	$-$0&14 		&	$-$1&39 	&	1&89 	&&	$-$0&04 		&	$-$0&65 	&	0&13 	\\
ADSI	&	$-$0&09 		&	$-$1&60 	&	0&74 	&&	0&04 		&	0&82 	&	0&13 	\\
KCFSI	&	0&12 		&	1&16 	&	1&32 	&&	0&05 		&	0&66 	&	0&27 	\\
Payroll Growth	&	$-$0&10 		&	$-$0&93 	&	1&05 	&&	$-$0&02 		&	$-$0&29 	&	0&04 	\\
SRP	&	0&20$^{**}$	&	2&17 	&	3&79 	&&	0&09 		&	1&34 	&	0&81 	\\
GDPG	&	$-$0&16$^{**}$	&	$-$2&55 	&	2&40 	&&	$-$0&13 		&	$-$1&64 	&	1&70 	\\[-3pt] 
\bottomrule

\end{tabular}
\end{table}

\clearpage \pagebreak

\end{document}